\documentclass[twocolumn]{aastex63}
\usepackage{makecell}

\graphicspath{{./}{figures/}}
\turnoffeditone
\shorttitle{Rotational Modulations of GU\,P\lowercase{sc\,b}}
\shortauthors{Lew et al.}

\begin{document}

\title{Cloud Atlas: Weak color modulations due to rotation in the  planetary-mass companion GU\,Psc\,b and 11 other brown dwarfs}

\correspondingauthor{Ben W.P. Lew}
\email{weipenglew@email.arizona.edu}
\setcounter{footnote}{0} 
\renewcommand*{\thefootnote}{\fnsymbol{footnote}}

\author[0000-0003-1487-6452]{Ben W.P. Lew}
\affil{Lunar and Planetary Laboratory, The University of Arizona, 1640 E. University Blvd,Tucson, AZ 85718, USA}
\author[0000-0003-3714-5855]{D\'aniel Apai}
  \affil{Department of Astronomy and Steward Observatory, The University of Arizona, 933 N. Cherry Ave., Tucson, AZ, 85721, USA}
  \affil{Lunar and Planetary Laboratory, The University of Arizona, 1640 E. University Blvd, Tucson, AZ 85718, USA}
  \affil{Earths in Other Solar Systems Team, NASA Nexus for Exoplanet System Science}
\author[0000-0003-2969-6040]{Yifan Zhou$^{\ast}$}\footnote{Harlan J. Smith McDonald Observatory Postdoctoral Fellow}
  \affil{Department of Astronomy, University of Texas, Austin, TX 78712, USA}
  \author[0000-0002-6732-3651]{Jacqueline Radigan}
\affil{Utah Valley University, 800 West University Parkway, Orem, UT 84058, USA}
 \author[0000-0002-5251-2943]{Mark Marley}
  \affil{NASA Ames Research Center, Naval Air Station, Moffett Field, Mountain View, CA 94035, USA}
  \author[0000-0002-4511-5966]{Glenn Schneider}
  \affil{Department of Astronomy and Steward Observatory, The University of Arizona, 933 N. Cherry Ave., Tucson, AZ, 85721, USA}
\author[0000-0001-6129-5699]{Nicolas B. Cowan}
  \affil{Department of Physics, McGill University, 3600 rue University, Montr\'eal, QC, H3A 2T8, Canada}
    \affil{Department of Earth \& Planetary Sciences, McGill University, 3450 rue University, Montr\'eal, QC, H3A 0E8, Canada}
\author[0000-0003-2446-8882]{Paulo A. Miles-P\'{a}ez$^{\dagger}$}\footnote{ESO Fellow}
\affil{European Southern Observatory, Karl-Schwarzschild-Stra{\ss}e 2, 85748 Garching, Germany.}
\author[0000-0003-0192-6887]{Elena Manjavacas}
  \affil{W.M. Keck Observatory, Mamalahoa Hwy, Kamuela, HI 96743, USA}
\author[0000-0001-7356-6652]{Theodora Karalidi}
  \affil{Department of Physics, University of Central Florida, 4000 Central Florida Boulevard, Orlando, FL 32816, USA}
\author[0000-0003-4080-6466]{L.\,R.\,Bedin}
  \affil{INAF-Osservatorio Astronomico di Padova, Vicolo dell'Osservatorio 5, I-35122 Padova, Italy}
\author[0000-0001-8014-0270]{Patrick J. Lowrance}
\affil{IPAC-Spitzer, MC 314-6, California Institute of Technology, Pasadena, CA 91125, USA}
\author{Adam J. Burgasser}
\affil{Astrophysics and Space Science, University of California San Diego, La Jolla, CA 92093, USA}

\renewcommand*{\thefootnote}{\arabic{footnote}}
\setcounter{footnote}{0}

\begin{abstract}
Among the greatest challenges in understanding ultra-cool brown dwarf and exoplanet atmospheres is the evolution of cloud structure as a function of temperature and gravity.
In this study, we present the rotational modulations of GU\,Psc\,b -- a rare mid-T spectral type planetary-mass companion at the end of the L/T spectral type transition.
Based on the HST/WFC3 1.1-1.67$\rm\, \mu m$ time-series spectra, we observe a quasi-sinusoidal light curve with a peak-to-trough flux variation of 2.7\% and a minimum period of eight hours.
The rotation-modulated spectral variations are weakly wavelength-dependent, or largely gray between 1.1-1.67$\rm\,\mu$m. The gray modulations indicate that heterogeneous clouds are present in the photosphere of this low-gravity mid-T dwarf. We place the color and brightness variations of GU\,Psc\,b in the context of rotational modulations reported for mid-L to late-T dwarfs. Based on these observations, we report a tentative trend: mid-to-late T dwarfs become slightly redder in $J-H$ color with increasing $J$-band brightness, while L dwarfs become slightly bluer with increasing brightness. If this trend is verified with more T-dwarf samples, it suggests that in addition to the mostly gray modulations, there is a second-order spectral-type dependence on the nature of rotational modulations.
\end{abstract}
\section{Introduction}

One of the most perplexing observations of the ultracool atmospheres of brown dwarfs and directly-imaged exoplanets is the prominent color evolution across the L/T spectral type transition. Over a narrow temperature range ($\sim$100~K) the atmospheres transition from red (in the near-infrared, $J-H \sim 1.3$) to blue ($J-H \sim 0.0$) colors. It has been proposed that this color evolution could be caused by cloud thinning \citep{ackerman2001,saumon2008}, cloud patchiness \citep{burgasser2002,marley2010}, cloud structure evolution \citep{tsuji2003, burrows2006, charnay2018}, and possibly even $CO/CH_4$ compositional-gradient driven instability \citep{tremblin2016,tremblin2019}. Brown dwarfs with mid-T spectral types also tend to be about 0.5 magnitudes brighter in the J-band than earlier and later spectral type counterparts -- also known as the J-band brightening, possibly as a result of cloud disruption \citep{burgasser2002,burgasser2006b,dupuy2012}. While models of ultracool atmospheres had considerable success in fitting the L spectral type sequence (with thick silicate condensate clouds) and the late-T spectral type brown dwarfs (with mostly cloud-free atmospheres), the rapid color evolution and brightening across the L/T transition point to the existence of processes not well understood.

Therefore, understanding atmospheric and cloud evolution from late-L spectral types though the L/T transition to late-T dwarfs remains an important challenge. 
It is clear that the color changes carry important information about the nature of the processes that occur in these cooling atmospheres.

Surface gravity may be part of the puzzle, too. Small samples of L/T transition brown dwarfs suggest that the L/T transition occurs at lower effective temperatures for low-gravity objects  \citep{metchev2006,dupuy2009,marley2012,bowler2013,liu2016,miles-paez2017}. The magnitude of $J$-band brightening could also be larger for low-gravity objects if we include directly imaged planets \citep[see Fig. 16 in][]{dupuy2012}. Alas, as cooling objects cross swiftly the L/T transition, very few brown dwarfs are known with low masses at the L/T transition, making it difficult to test model predictions about the interplay of surface gravity and cloud evolution.

A particularly important probe of atmospheric properties is time-resolved high-precision (sub-percent level) spectrophotometry, that can -- through the rotation of the target -- explore non-uniform brightness distribution in an atmosphere with fixed gravity and interior temperature. Such rotational mapping studies have been used successfully to constrain cloud properties in ultracool atmospheres, including those with planetary masses \citep[e.g.,][]{apai2013,buenzli2014,metchev2015,zhou2016,biller2017,apai2017,manjavacas2019b,miles-paez2019,zhou2019}.
The wavelength dependence of rotation modulations sheds light on the variations of cloud particle sizes, molecular abundances, and photospheric temperatures. In a rotating atmosphere, the temporal modulations at a given wavelength probe the atmosphere's spatial structure in a pressure range. 
Consequently, comparisons of the modulations observed at different wavelengths probe pressure-dependent properties in the atmosphere.

Multiple studies have used time-resolved ground-based photometry \citep[e.g.,][]{artigau2009,radigan2012,biller2013,radigan2014a}, spectroscopy \citep[e.g.,][]{schlawin2017}, or space-based spectrophotometry \citep[e.g.,][]{buenzli2012,buenzli2015a,yang2015,karalidi2016,lew2016,yang2016} to explore the variations of the near-infrared colors of rotating brown dwarfs (from mid-L to T8 spectral types). All of these studies found gray, i.e., only weakly wavelength-dependent modulations in the near-infrared, even for objects with large-amplitude ($>$10\%) modulations.
Radiative-transfer-based models presented in \citet[][]{apai2013} explain the gray modulations with a correlated change in effective temperature and cloud thickness; this study and that of \citet{radigan2012} show that changes in a single model parameter (temperature {\em or} cloud thickness) cannot explain the observed modulations. 
Similar conclusions have been drawn by \citet{biller2018}, who observed the planetary-mass late-L dwarf PSO J318.5338-22.8603.
They suggest that the heterogeneous high-altitude clouds or extended silicate clouds could explain the weak wavelength-dependence of near-IR and mid-IR modulation amplitudes, as well as the phase offset between the near-IR and mid-IR light curves.

These examples demonstrate the power of time-resolved high-precision spectrophotometry in constraining the heterogeneous cloud properties of individual atmospheres and -- through comparisons of objects spanning the L--L/T--T sequence -- the potential for deciphering cloud evolution in cooling atmospheres. For a more complete discussion of results from time-resolved studies we refer to recent reviews \citep{biller2017,artigau2018}.

In this paper we present a new, space-based and high-precision time-resolved spectrophotometric study of \object{GU\,Psc\,b}, one of the rare planetary-mass brown dwarfs at the end of the L/T sequence. In Section~\ref{s:Target} and \ref{s:Observations} we described the target, the observations and data reduction process. We present the spectra and light curves in Section~\ref{s:Results}. \edit1{In Section~\ref{s:Regression}, we describe the compilation of published data and the analysis of the color-magnitude variations of twelve brown dwarfs}.  We discuss the implications of our results to the color change in the L/T transition in Section \ref{s:Discussion} and summarize our conclusions in Section~\ref{s:Conclusions}.

\section{GU P\lowercase{sc} and GU P\lowercase{sc b}}
\label{s:Target}
GU Psc (or \object{2MASS J01123504+1703557}) is an M3.5 dwarf at $47.6\rm \,pc$ based on the Gaia parallax of $21.00 \pm 0.07\rm \, mas$ \citep{GaiaCollaboration2016,GaiaCollaboration2018,luri2018}.
Based on its kinematic and photometric properties, \citet{malo2013} categorized the GU Psc system as a highly probable (96.9\%) member of AB Doradus moving group (ABDMG), which is around $120 \pm 30$ Myr old \citep{zuckerman2004}.
The detection and measured width of $H\alpha $ emission of GU Psc by \citet{riaz2006} indicates that the age of the system is in between $10\rm\,Myr$--$\rm2\,Gyr$ respectively \citep{barrado2003,white2003,west2008}.
\citet{naud2014}, hereafter N14, find that the calculated X-ray luminosity ($\rm \log(L_x) = 29.1 \pm 0.3\,ergs^{-1}$ at $48 \pm 5\rm\, pc$) of GU\,Psc based on the ROSAT observation is similar to that of other single M dwarfs in ABDMG and is higher than that of field stars.
The kinematic properties, $\rm H\alpha $ emission, and X-ray luminosity measurements together suggest that GU\,Psc is a relatively young system compared to field dwarfs.

N14 provides an estimate of the GU Psc's metallicity ([Fe/H]) that ranges from ${-0.14} \pm 0.09$ to $0.1 \pm 0.13$ with various methods \citep{Mann2013,Newton2014}.
The measured periodic variability of $1.0362 \pm 0.0005$ days \citep{norton2007} and $v \sin i$ ($23\, \rm km\,s^{-1}$, N14) of GU Psc suggests that it is a rapid rotator.
More detailed characterization of GU Psc system can be found in N14.

\object[GU Psc b]{GU\,Psc\,b}, discovered by N14, is at a projected distance of $2,000 \pm 200$ au from GU Psc.
Based on the near-infrared spectral index and the comparison with standard T dwarfs, N14 classifies GU\,Psc\,b as a T3.5$ \pm 1$ dwarf.
N14 finds the best-fit effective temperature ranges from 1,000 to 1,100\,K by comparing the GU Psc b's near-infrared spectrum with atmospheric models \citep{baraffe2003,saumon2008}.
Based on the ABDMG's age of $100 \pm 30$\,Myr and the fitted effective temperature range,
N14 estimates the mass of GU\,Psc\,b to be around 9-13 $M_{\rm  Jup}$, which is close to the deuterium-burning mass limit of 12-13 $\rm M_{\rm Jup}$ \citep{saumon2008} -- commonly adopted as the borderline between brown dwarfs and giant planets with solar metallicity.
\citet{naud2017} report a tentative J-band photometric variability of GU\,Psc\,b with a peak-to-trough flux variation of $4 \pm 1\%$ at a period of 5.9 hours based on one out of three nights of 5-6 hour observation with WIRCam Imager at 3.6m Canada-France-Hawaii Telescope.

\section{Observation and Data Reduction}
\label{s:Observations}

\begin{figure*}
    \centering
    \includegraphics[width=.9\textwidth]{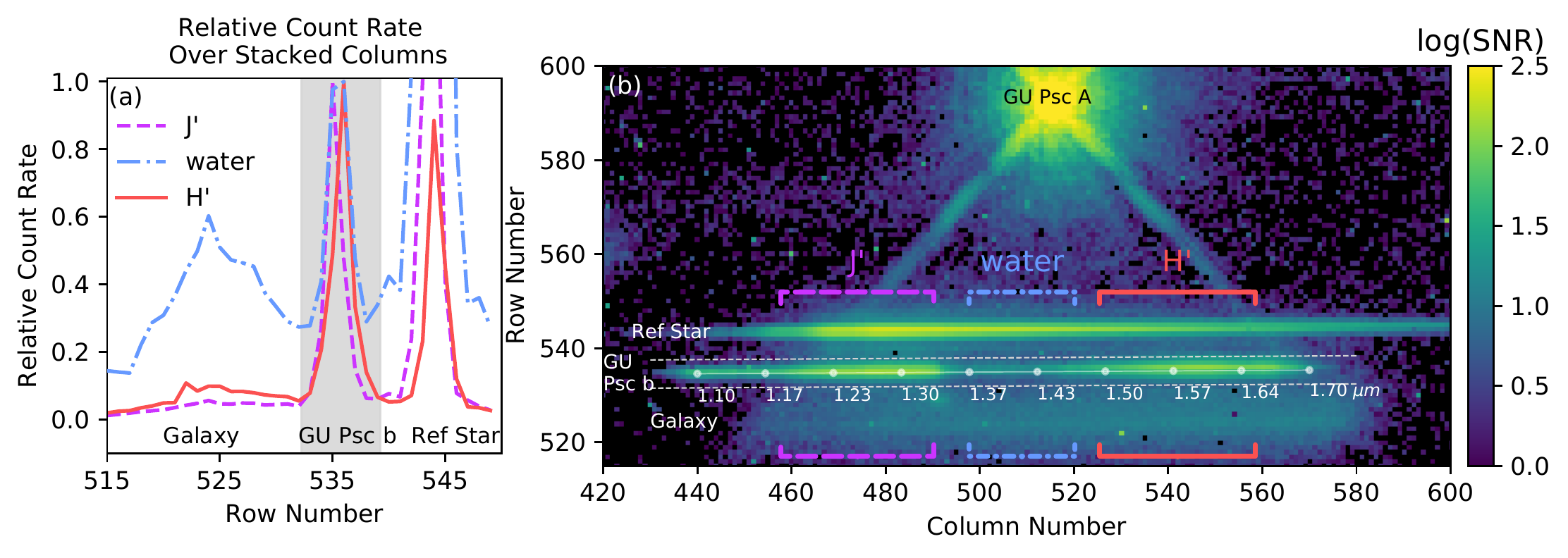}
    \caption{
    (a): The relative count rates of pixels stacked along the rows in panel (b) in three different wavelength regions: approximated J band (dashed magenta line), water band (dashed-dot blue line), and H band (solid red line). In the spectral extraction region of GU\,Psc\,b, plotted in shaded gray, the flux contribution from nearby sources is lower than $\sim 10\%$ for $J'$ and $H'$ bands, but higher than $\sim 20\%$ for water band.
    (b) A cropped median image to illustrate the nearby reference star and galaxy spectra flanking that of GU\,Psc\,b. The image is color-coded with signal-to-noise ratio in log scale. The three square brackets represent the $J'$-, water-, and $H'$-band regions in which the pixel count rates are stacked and summed along the rows. The dot-connected line with annotated wavelengths in microns is plotted for GU\,Psc\,b's spectrum. The two white dashed lines that bracket the dot-connected line mark the aperture (six pixels) for spectral extraction. }
    \label{fig:fov}
\end{figure*}
 We utilized six consecutive Hubble Space Telescope (HST) orbits to observe the rotational modulation of GU\,Psc\,b in Wide Field Camera 3 (WFC3)'s G141 grism mode ($1.07-1.70\rm\,\mu m$, spectral resolving power $\sim 130$ or dispersion of 4.7 nm/pixel with a plate scale of 0.13"/pixel) on Jan 08 2018.
These observations are part of the HST Large Treasury Program Cloud Atlas (P.I: Apai, ID: GO-14241). There are eleven 201\,s-long integration spectroscopic frames in each 96-minute long HST orbit.
We restricted HST's orientation angle to minimize possible spectral contamination from galaxies and bright stars within small angular separations from GU\,Psc\,b.

The data reduction pipeline is similar to that in \citet[][]{lew2016}. In brief, our data reduction process started from {\tt{flt.fits}} files, which are produced by the {\tt{calwf3}} pipeline for zero-read and bias calibration, dark image subtraction, flat-fielding, detector nonlinearity correction, and cosmic rays flagging. Our well-vetted pipeline \citep[e.g.,][]{buenzli2012,apai2013,yang2015,zhou2018} then interpolated around the cosmic-rays affected and other bad pixels before subtracting the sky background.  We followed the method in \citet{kummel2011} for background reduction that scales the master sky image from \citet{kummel2011} for background reduction.
We used \texttt{Source Extractor} \citep{bertin1996} to extract source location from the stacked direct images that are observed with F132N filter at the beginning of each orbit for wavelength calibration.
We used {\tt{aXe}} \citep{kummel2009} with a six-pixel wide cross-dispersion aperture for spectral extraction.
We performed an absolute flux correction for finite aperture photometry by interpolating the table of encircled energy as a function wavelength and diameter of Table 6b in \citet{kuntschner2011}.

We did not find a significant ramp effect \citep{smith2008,long2014} in the six-orbit-long observation, therefore no ramp correction \citep[e.g.,][]{zhou2017} was performed. The less pronounced ramp effect in our data is likely because of the low count rate (peak count rate $< 8 e^{-}s^{-1}$ with an averaged count rate \footnote{The averaged count rate was calculated over a box that is six-pixel wide in cross-dispersion direction and 130-pixel long in the spectral dispersion direction of GU\,Psc\,b.} of $\sim 2.8 e^{-}s^{-1}$) compared to that of the case studies in \citet{zhou2017}. 
Given a lower count rate, the ramp profile becomes more linear and is less significant compared to the photon noise. 
Any uncorrected ramp effect, which mostly increases the flux in the first orbit, will only increase the brightness variations reported in this study. 

\subsection{Contamination assessment}\label{sec:contam}
As Figure~\ref{fig:fov} shows, there are two sources in close angular proximity of GU\,Psc\,b, leading to a second-order spectrum of a reference star and to a first-order spectrum of a galaxy (see also Figure 1 in N14). Also, there is a faint diffraction spike from the GU Psc's 0$^{th}$-order (undispersed) grism image superimposes upon part of  GU\,Psc\,b's spectrum. To mitigate this, we interpolate the flux density in the 1.17-1.19$\rm\, \mu m$ region of the GU Psc b spectra (see Figure \ref{fig:spec}) to avoid possible contamination from GU Psc's 0$^{th}$-order diffraction spike.

To assess possible contamination from the two nearby sources, we sum the measured count rates in the same row (x-axis) over three ranges of columns (i.e., the bracketed regions colored in magenta, blue, and red in Figure \ref{fig:fov}b), approximating the measured count rate in the $J'$ (1.18-1.33$\rm\, \mu m)$, water (1.37-1.47$\rm\, \mu m)$, and $H'$ bands (1.50-1.65$\rm\, \mu m)$.
The summed count rates is then plotted as a function of row number in Figure~\ref{fig:fov}a. In Figure~\ref{fig:fov}a, the GU\,Psc\,b's summed count rate spans roughly across rows 532$-$538 and is highlighted by the gray band.
The gray region is similar to the aperture used for spectral extraction\footnote{The direction of spectral trace and the corresponding aperture for spectral extraction are not perfectly aligned with the x-axis of the image grid but a field-dependent multi-order polynomial of x and y axes \citep{kuntschner2009}. Therefore, the row number at where the GU Psc b's count rate reaches the maximum in the J' and H' bands are slightly different.}.
We normalize the summed count rate of the three ranges of column so that the GU Psc's peak count rate equals one.

 Figure~\ref{fig:fov}a provides an order-of-magnitude estimation of the contamination level -- within the spectral aperture for GU\,Psc\,b, the contamination level is less than 10\% of the GU\,Psc\,b's peak count rate in the $J'$ and $H'$ bands, but much higher ($>20\%$) in the water band.
 A more sophisticated analysis that fits 1D 3-Moffat profiles to the summed count rate from 1.1-1.7$\rm \,\mu m $ (column number 440-570) gives a similar estimate of contamination level (13\%, see also Figure \ref{fig:contam}).
 Because of the low signal-to-noise ratio and the moderate level of contamination in the water band,  in this study we choose to focus on the $J'$-band, $H'$-band, and the integrated white ($1.1-1.67\rm\, \mu m$) light curve.

 In Figure \ref{fig:sinfit}, the reference star's light curve shows a slight brightening trend at a sub-percent level. A simple straight-line fit to the reference star's light curve gives a slope of $(9 \pm 2) \times 10^{-4}$/hour, or 0.7\% for an eight-hour baseline. This linear-brightening trend is possibly an HST systematics that leads to visit-long slopes \citep[e.g.,][]{berta2012,wakeford2016}. In the field of view no other source's light curve has similar signal-to-noise ratios to verify the presence of this possible sub-percent-level systematics. The low signal-to-noise ratio light curve of the nearby background galaxy fluctuates at about four-percent from the brightest to the dimmest state.  
 The maximum contamination level of flux variation from the nearby sources is roughly equal to the product of flux contamination fraction and the flux variation of nearby sources, which is about $13\% \times 4\% = 0.52\%$ level. The estimate of 0.52\%, which is a generous upper limit considering the different light curve profiles of between GU\,Psc\,b and other sources, is much lower than the observed peak-to-trough flux variation of 2.7\% (see Section \ref{s:amplitude}) in the integrated white light curve of GU\,Psc\,b. 

 We also check the distribution of flux density deviations from the median spectrum. 
 We find only two data points with 3.2-$\sigma$ deviation from the median spectral flux densities among 124 spectral bins in 66 exposures. Therefore, no statistically significant deviation from the median spectrum is found assuming spectral points are independent of each other.
 Therefore, our careful inspection of spectral variations and reduced images confirms that the observed rotational modulation is intrinsic to GU\,Psc\,b.

\section{Spectra and Rotational Modulations}
\label{s:Results}

\subsection{Spectrum and spectral variations}
\begin{figure}
    \centering
    \includegraphics[width=.5\textwidth]{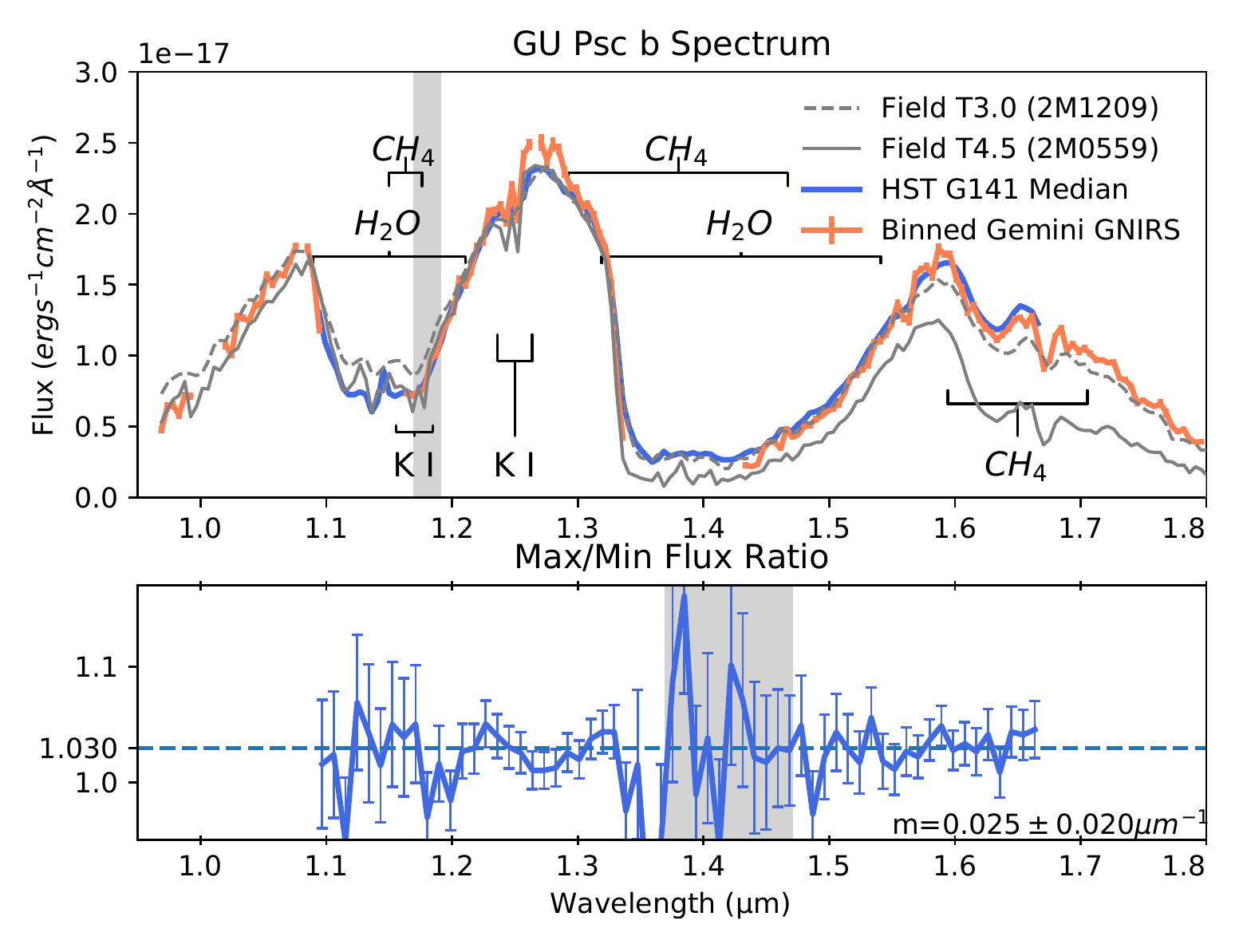}
    \caption{(a) The median-averaged spectra plotted in blue in comparison with the GNIRS spectrum in solid orange line. Spectra of field T4.5 dwarf 2M0559 (solid gray line) and T3.5 2M1209 (dashed gray line), which are scaled to match the J-band maximum flux density of GU\,Psc\,b, are also plotted for comparison.\edit1{The wavelength region in which the flux density is interpolated is colored in grey.}(b) The max/min flux ratio among the six HST-orbit averaged spectra suggests no strong wavelength dependence in the rotational modulations. The fitted wavelength-dependence slope $m$, excluding the water-band (gray region), is shown at the bottom right.}
    \label{fig:spec}
\end{figure}

 We plot the HST median spectra together with the Gemini Near-InfraRed Spectrograph (GNIRS) spectrum (R$\sim$800) from N14 in the top panel of Figure \ref{fig:spec}.
We also overplot the spectra of two field T dwarfs, T4.5 \object{2MASS J05591914$-$1404488} \citep{burgasser2006a} and T3 \object{2MASS J12095613$-$1004008}
 \citep{burgasser2004} that are normalized to the same J-band peak flux density as that of GU\,Psc\,b.
Our HST/G141 observations provide the first flux density measurements of GU\,Psc\,b in the water-band absorption region of 1.1-1.2$\rm\,\mu m$. After the J-band flux normalization of the two field T dwarfs' spectra, GU Psc b's spectrum matches better the T4.5 spectrum at wavelengths $\lambda <1.3\rm\, \mu m$ but matches better the T3 spectrum at longer wavelengths ($\lambda >1.3\rm\, \mu m$).

In the bottom panel of Figure \ref{fig:spec}, we plot the binned ratio of the median spectra of the sixth and the third orbit, which are the orbits at which the broadband-integrated flux density reaches its maximum and minimum. The max/min flux ratio shows no strong wavelength-dependence for the rotational modulations, similar to that of T0 dwarf SIMP 0136 \citep{apai2013}. After excluding the low-signal-to-noise-ratio water band (1.37-1.47$\rm\,\mu m$), the mean max/min flux ratio is about 3.0\%. The fitted slope (m =$ 0.025 \pm 0.020\rm \,\mu m^{-1}$) for the max/min flux ratio suggests that there is no significant wavelength-dependence.

\subsection{Light curves of rotational modulation} \label{s:amplitude}

\begin{figure}
    \centering
    \includegraphics[width=0.47\textwidth]{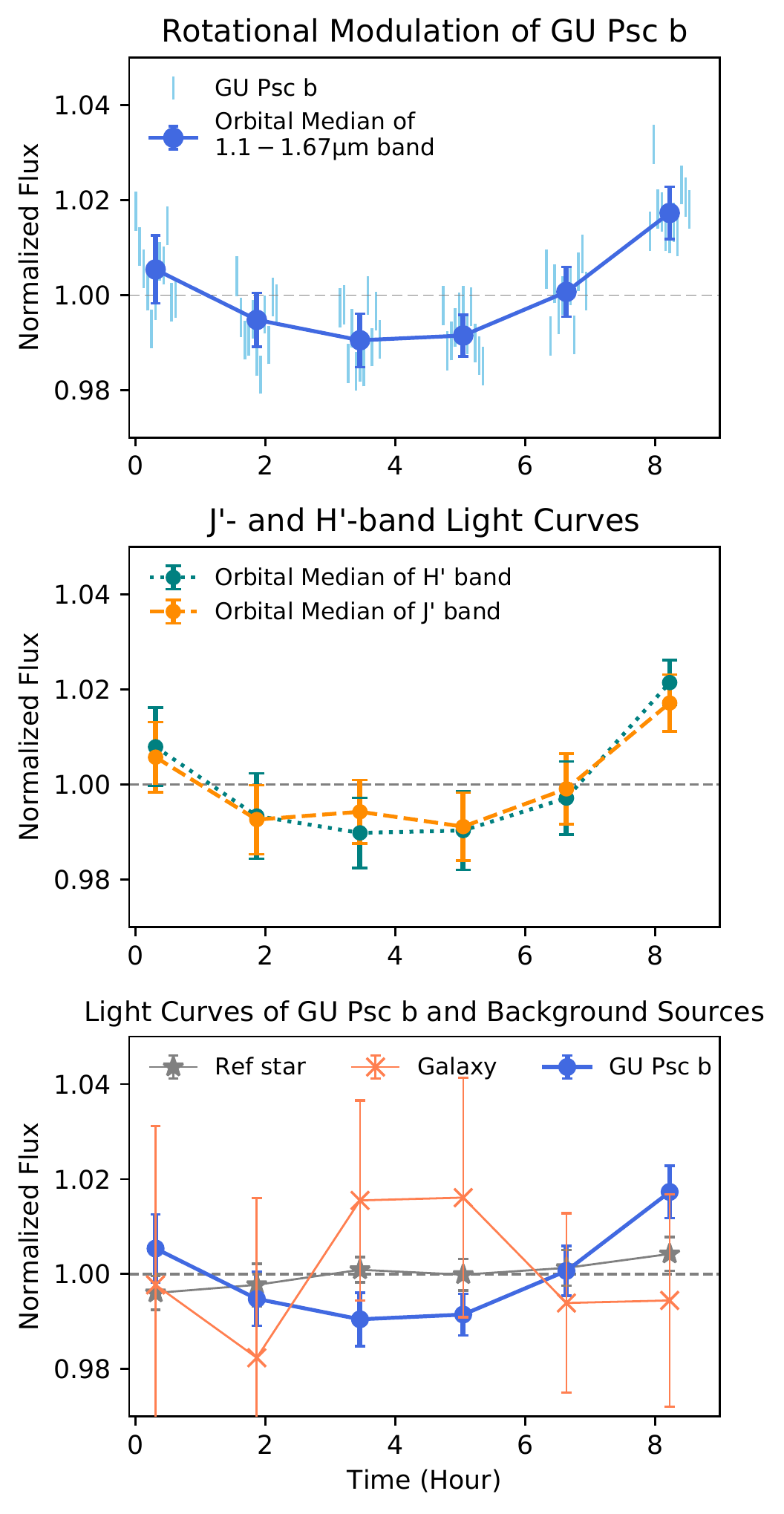}
    \caption{Top Panel: The rotational modulation of GU\,Psc\,b shows a $2.7 \pm 0.8 \%$ peak-to-trough flux variation during the eight-hour observation. \edit1{In each HST orbit, we plot eleven single-exposure photometric points as light blue error bars and the orbital median as dark blue points. The error bar of orbital median is the standard deviation of photometric points per HST orbit}. Middle Panel: The $J'$- and $H'$-band light curves show similar light curve profiles. Bottom Panel: GU\,Psc\,b's white light curve is distinct from the flux variations of the nearby background sources.}
    \label{fig:sinfit}
\end{figure}

We plot the integrated white (1.1-1.67$\rm\,\mu m$) light curves of GU\,Psc\,b and those of the other two nearby sources in Figure \ref{fig:sinfit}. The GU\,Psc\,b's light curve manifests a sinusoidal profile with a period longer than the observation baseline. The sinusoidal pattern of GU\,Psc\,b's light curve is distinct from the almost flat light curve of the reference star and the choppy light curve of the nearby galaxy.
The $J'$-band ($1.18-1.33\rm\,\mu m$) and $H'$-band ($1.50-1.65\rm\,\mu m$) light curves also show a similar profile as that of the integrated white light curve.
 Based on the integrated white light curve, the rotational modulation amplitude is at least 1.35\%, or $2.7 \pm 0.8 \%$ for peak-to-trough flux variation \edit1{(i.e., the ratio of the integrated flux median at the 6th orbit to that at the 3rd orbit)}. This variability level is consistent with the previously reported marginal detection of peak-to-trough variability of $4 \pm 1\%$ at a timescale of $\sim 6\,$hour by \citet{naud2017}.
 
 \edit1{Because of the incomplete phase coverage, the fitted rotational period is degenerate with the amplitude for a sinusoidal model (c.f., Figure \ref{fig:mcmc} for the posterior distribution of Markov Chain Monte Carlo fitting result). Therefore, we only place a lower limit of eight hours on the rotational period, corresponding to the baseline of the HST observations. 
}

\section{Rotational Modulations on the Color-Magnitude Diagram}\label{s:Regression}

Color-magnitude diagram (CMD) is a useful tool for studying the brown dwarf atmosphere evolution with thousands of brown dwarfs photometric and parallax observations \citep[e.g.,][]{dupuy2012,best2015}. 
Meanwhile, an increasing number of brown dwarfs with detected rotational modulations through HST/G141 time-series spectral observations have been reported.
\edit1{We compile eleven brown dwarfs with published HST/G141 spectral observations (\object{2MASS J22282889$-$4310262} \citealt{buenzli2012}; \object{SIMP J013656.5+093347.3} \& \object{2MASS J21392676+0220226} \citealt{apai2013}; \object{Luhman 16B} \citealt{buenzli2015a}; \object{2MASS J15074769-1627386} \& \object{2MASS J18212815+1414010} \citealt{yang2015}; \object{WISEP J004701.06+680352.1} \citealt{lew2016}; \object{HN Peg B} \citealt{zhou2018}; \object{PSO J318.5338$-$22.8603} \citealt{biller2018}; \object{LP 261$-$75 B} \citealt{manjavacas2018}; \object{Ross458c} \citealt{manjavacas2019b}) to study the color dependence of rotational modulations across different spectral types in the 2MASS $M_J$ vs $J-H$ CMD. In this section, we describe the conversion from HST spectra to broadband photometry, the empirical models for color-magnitude variations, and discuss the overall trend in color-magnitude variations on CMD.}

\subsection{Binning HST time-series spectra to broadband photometries}\label{sec:bin}

The HST/G141 grism's wavelength coverage does not fully overlap with that of the 2MASS $H$ band.
By weighting the HST/G141 spectral variations in the HST $J'$ (1.18-1.33$\rm\, \mu m$) and $H'$ (1.50-1.65$\rm\, \mu m$) bands with the 2MASS $J$-band and H-band spectral response curves \citep{wright2010}, we implicitly assume that modulation amplitudes $\Delta J' = {\rm2MASS}\,\Delta J$ and $\Delta H' = {\rm2MASS}\, \Delta H$.

To plot the HST/G141 spectral variations on a 2MASS CMD, we also adopt the 2MASS $J$- and $H$-band magnitudes \citep{cutri2003} as the mean magnitudes of the HST $J'$- and $H'$-band modulations, except for the most variable object 2M2139\footnote{Even with the assumption of $\Delta J' = {\rm2MASS}\,\Delta J$, the mean of HST $J'(t)$ does not necessarily equal to 2MASS $J$, especially for object with large modulation amplitude.}. We first scale the $J$-band peak of the 0.6-2.65$\rm\, \mu m$ 2M2139 spectrum from the SpeX library \citep{burgasser2014a} to be the same as that of the averaged HST spectra. Then we use the scaled SpeX spectrum to calculate the mean magnitude in the 2MASS $J$ and $H$ bands during the HST observation of 2M2139.

\edit1{The timescale of intrinsic color-magnitude variations from rotational modulations of brown dwarfs is typically on the order of hours. 
Meanwhile, the observed color-magnitude variations on several minutes timescale are likely dominated by photon noise and/or systematics.
To study the intrinsic color-magnitude variations across the rotational phase, for objects with long rotational periods (P$ > $5 hours) we bin the photometric points in the each HST orbit (total exposure time of $\sim $ 30 to 40 minutes) with the median value.
We estimate the uncertainties of the median with the standard deviations of color and magnitude variations in each bin.
These uncertainties are conservative estimates because they include the photon noise and the readout noise, variations from time-variable systematics (e.g., ramp effect), and intrinsic variability of objects.
For objects with periods less than 5 hours, we choose not to bin the photometric points so that the cadence of color-magnitude variations is less than 10\% of the rotational phase. }

\subsection{Empirical models for color-magnitude variations}
The $J-H$ color change due to rotational modulations in any individual object is much smaller compared to the $J-H$ color evolution across the L/T transition.
To visualize the small scale of color change and the large scale of color evolution in CMD, we fit a straight line to the magnitude-color variation (i.e. $M_J$ vs. $J-H$) for each object's rotational modulations.
We then plot the fitted straight line of magnitude-color variation on CMD.
We obtain the best-fit slope and y-intercept using an orthogonal distance regression algorithm with {\tt{scipy.odr}} (see also Figure \ref{fig:odr}), which minimizes the orthogonal distance, weighted by both x- and y-axis uncertainties, between photometric points and the straight-line model. 
\edit1{We use the covariance matrix of the fitted parameters to calculate the standard deviation of the fitted slope and y-intercept.}

We remind that this linear fit is only for illustrating the primary direction of color changes with respect to the brightness variations. The color-magnitude variations could be non-linear, too. 
\edit1{
For example, the subpanel plots in Figure \ref{fig:cmd} suggest that a linear model may not fit well to the color-magnitude variations of GU Psc b and 2M2228.
As one of the simple, periodic non-linear models, we fit an ellipse to the color-magnitude variations and show the direction of color change direction with the semi-major axis of the ellipse.
The fitted color-magnitude changes of ellipse are not unique (e.g., the elliptical and circular fit for GU\,Psc\,b in Figure \ref{fig:cmd}) and is sensitive to the phase coverage of light curves.
}

For reference, we also plot the color-magnitude variations of a blackbody with varying temperatures, either by hotter or cooler from the effective temperature for each object as a dashed gray line in the subpanels of Figure \ref{fig:cmd}. In the near-infrared $J$ and $H$ bands, a sum of blackbodies with different temperatures is still similar to a Planck function \citep{schwartz2015}. Therefore, this model acts as a toy model of a heterogeneous photosphere with a mixture of blackbodies at different temperatures.

\subsection{Result of fitting magnitude-color variations}\label{sub:cmdresult}
\begin{figure*}
    \centering
    \includegraphics[width=1.0\textwidth]{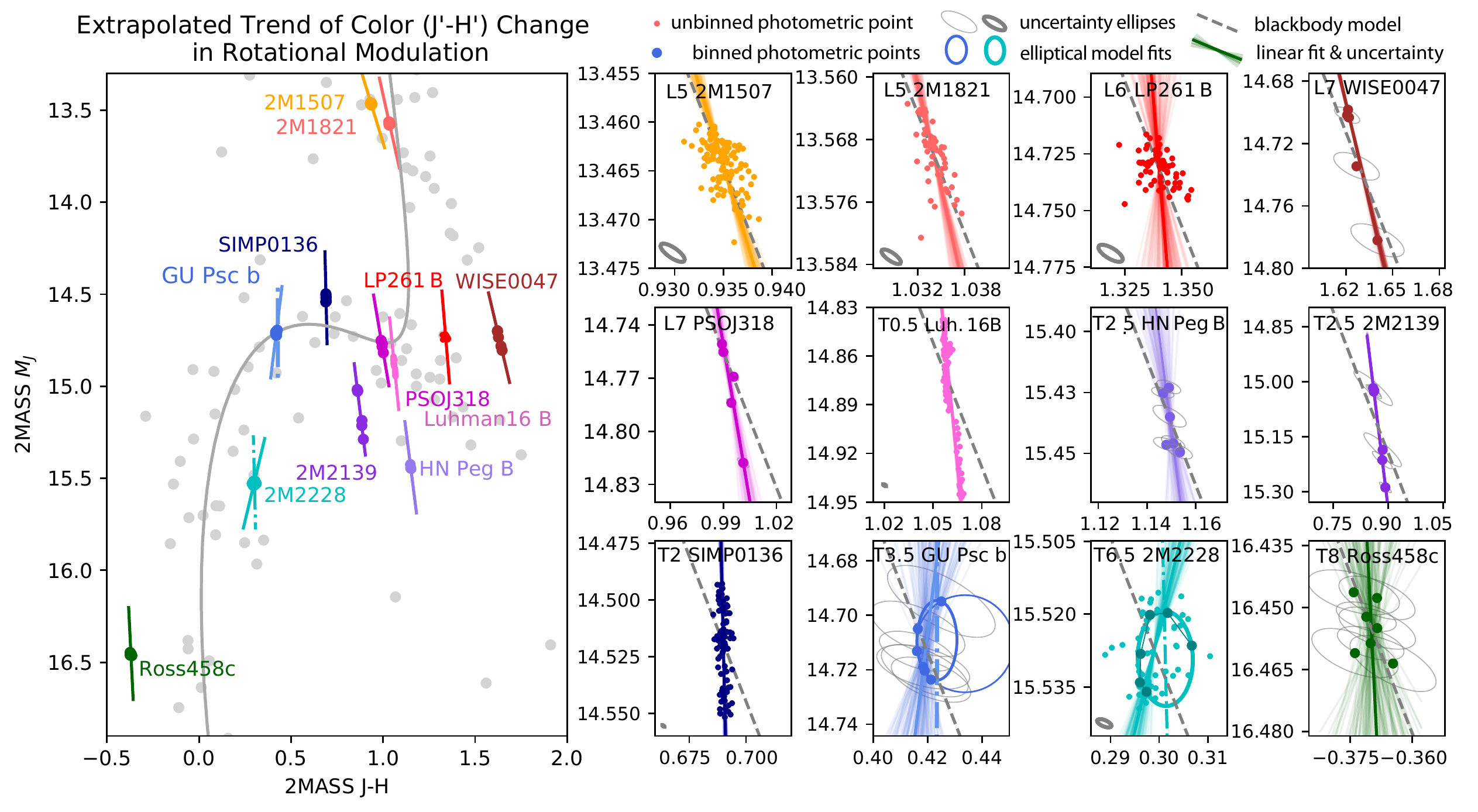}
    \caption{{\bf{Left panel}}: \edit1{The fitted linear trajectories of color-magnitude variations for twelve objects are plotted in solid lines. 
    The almost vertical direction of the plotted trajectories suggests that most objects show relatively weak color changes compared to their rotational modulation amplitudes. Dash-dotted lines are the semi-major axes of the fitted ellipses. 
    The gray dots represent the colors and magnitudes of brown dwarfs from \citet{dupuy2012}; The gray line is the empirically derived color-magnitude evolution curve from \citet{dupuy2012}.
    {\bf{$\mathbf{3 \times 4}$ panels on the right}}: Zoomed-in plots for magnitude-color variations and the fitted slopes for individual objects. 
    We plot the HST-orbital photometric points (large solid dots) and the unbinned photometric points (small solid dots) for objects with long and short rotational periods respectively (c.f., Section \ref{sec:bin}).
    The uncertainty ellipses are plotted in grey. 
    For clarity, we plot the typical uncertainty ellipses of single-exposure photometric points at the bottom left corner of subpanels.
    Three models to fit the color-magnitude variations are plotted: straight line (solid straight lines), ellipse (for 2M2228 and GU\,Psc\,b, curved solid lines), and blackbody with varying temperature (gray dashed lines). See Section \ref{s:Regression} for more details of the models.
    The uncertainties of the linear models are plotted in semi-transparent color lines.
    The semi-major axes of fitted ellipses are plotted in dash-dotted lines.
    The left and right panels share the same aspect ratio. }
    }
    \label{fig:cmd}
\end{figure*}

\deleted{With the detected rotational modulation of T3.5 GU\,Psc\,b, we now have a small number of valuable objects (GU\,Psc\,b, Luhman 16B, HN Peg B, 2M2139, and SIMP0136) with detected simultaneous $J$-and-$H$-band modulations across the L/T transition ($\sim$ L8-T5).}
In the left panel of Figure~\ref{fig:cmd}, we present the trajectories of the rotational modulations of twelve ultracool atmospheres spanning a broad spectral type range. 
We plot the colors and magnitudes of field brown dwarfs whose with SNR$>$10 for $J$ and $H$ band photometry from the updated catalog of \citet{dupuy2012}. The solid gray line is the empirically-derived polynomial function of magnitude vs. spectral type from \citet{dupuy2012}.
\edit1{In the $3\times 4$ subpanels in Figure \ref{fig:cmd}, we show the color-magnitude variations of individual objects and the model fitting results.
We plot the linear fit and the blackbody models for every object.
For GU\,Psc\,b and 2M2228 that demonstrate non-linear color-magnitude variations, we also plot the best-fit ellipses for reference.
For 2M2228, the non-linearity of variations is more apparent with the HST-orbital photometric points.}

The slope of the fitted lines qualitatively shows how much the $J-H$ colors change with respect to the $J$-band flux variations.
Among the plotted objects, 2M2139 demonstrates the largest color change ($\Delta (J'-H') = 0.0378 \pm 0.0015$), but it is still much smaller than its $J'$-band magnitude variations ($\Delta J' \sim 0.300 \pm 0.001$).
All of the plotted objects show less change in their colors compared to their modulation amplitudes.
In other words, all objects show only weak color changes in their rotational modulations.

In the CMD, a positive slope suggests that the object becomes brighter and redder, and a negative slope suggests that the object becomes brighter and bluer.
Most of the objects show  negative slopes, especially among L dwarfs.
Among the plotted objects, only GU\,Psc\,b and 2M2228 show positive slopes.
Our fitted slope uncertainty distribution suggests that there is about 20\% probability for GU\,Psc\,b to have a negative slope as seen in other L dwarfs.
Therefore, the fitted slope of GU Psc b's color-magnitude variations is not significantly different from that of the other L/T transition objects.
There are two possible interpretations for the possible positive slope -- either there is a phase shift between $J'$- and $H'$-band light curves of GU\,Psc\,b, as in the case of 2M2228 (phase shift of $15 \pm 2 ^{\circ}$, \citealt{buenzli2012}), or the $H$-band modulation amplitude is indeed higher than that of $J$-band.
A longer baseline observation is needed to understand and verify if the color change of GU\,Psc\,b is indeed distinct from that of other L dwarfs.

\section{Discussions}
\label{s:Discussion}

\subsection{The first planetary-mass object with confirmed modulations at the end of the L/T transition}
The modulations of GU\,Psc\,b described here are the first in a planetary-mass object at the end of the L/T transition (T3-T5), confirming the previous $J$-band marginal detection by \citet{naud2017}.
Based on the observed rotational modulations, we also argue that low-gravity mid-T dwarf's photosphere is not cloud-free as -- with the spectral evolution from L to T type -- the silicate cloud base presumably sank to deeper pressures (see the condensation curves in \citealt{robinson2014,helling2014}).

In addition to the modulations observed in GU\,Psc\,b, we have gathered a unique data set of rotational modulations found in planetary-mass objects across different effective temperatures, including the L7 spectral type PSOJ318 \citet{biller2015,biller2018}, T2 type SIMP0136 \citep{apai2013}, and the T8 type Ross 458c \citep{manjavacas2019b}. Our dataset provides a useful reference point for future studies of cloud evolution across the L/T transition. Furthermore, the wavelength-dependence in rotational modulations of GU\,Psc\,b will be useful in testing predictions of different cloud models on the role of gravity in shaping cloud structure, as well as on testing the hypothesis that the L/T transition, and perhaps the $J$-band brightening too, depend on surface gravity.

\subsection{The modulation amplitude and rotational period}
In any rotating atmosphere, an asymmetric brightness distribution leads to rotational modulations. The modulation amplitudes cannot be directly translated to brightness maps, as the hemisphere-integrated measurements necessarily result in information loss (although some inferences can be drawn, see \citealt{cowan2008,apai2013,cowan2013,karalidi2016}).
The high modulation amplitude observed in GU\,Psc\,b suggests that there are prominent rotationally asymmetric features (dark or bright) that significantly ($>2\%$) impact even the hemisphere-integrated brightness of the photosphere. The features may be large and/or high contrast; if there are many features, these must be distributed asymmetrically.
From infrared rotational modulation surveys that include both high- and low-gravity brown dwarfs, we already know that brown dwarf atmospheres are ubiquitously heterogeneous \citep[e.g.,][]{buenzli2014,metchev2015}.
Typical field brown dwarfs outside the L/T spectral type transition have peak-to-trough flux variations of less than 2\% level in the $J$ band \citep{buenzli2014,radigan2014b}.
The observed peak-to-trough flux variations of $\sim$2.7\% of GU\,Psc\,b is consistent with \citet{radigan2014b}'s conclusion that L/T transition objects are likely to have the largest modulation amplitudes.
The high modulation amplitude of this low-gravity object provides another data point to test whether high modulation amplitude is correlated with low gravity, as claimed by \citet{vos2017} based on a compilation of published variability amplitudes of brown dwarfs.

\deleted{Assuming that the derived period of the quasi-sinusoidal modulations observed in the light curve of GU\,Psc\,b is similar to its intrinsic rotational period (see \citealt{apai2017}), the rotation period is at least eight hours.} The minimum period of eight hours is in line with the measured timescale of periodic modulations of other brown dwarfs, ranging from as short as 1.4 hours (\citealt{buenzli2012}) to 18 hours or longer (e.g. 2M2148 \citealt{metchev2015}).
 The actual rotational modulation profile may evolve with time, as seen in long-baseline observations of multiple L/T transition brown dwarfs \citep[e.g.,][]{apai2017}, most prominently detected in all three brown dwarfs (2M2139, 2M1324, and SIMP0136) monitored by the Spitzer Space Telescope in the Extrasolar Storms program \citep{yang2016,apai2017}. That study shows that the light curve evolution is the likely result of planetary-scale waves that modulated surface brightness \citep{apai2017}, possibly through the interplay of atmospheric circulations, condensations, and cloud formation/dispersal \citep{tan2017,tan2018,showman2018}. These mechanisms may also be present in GU\,Psc\,b and their presence could be revealed by continuous observations over 3-4 rotational periods.

The lower limit of the rotational period of GU\,Psc\,b also provides another data point for studying the evolution of spin as a function of mass and age. 
Assuming that the radius of GU\,Psc\,b is roughly 1.3-1.4 Jupiter radius, as expected for a 100\,Myr old and $10-12\,\rm M_{\rm Jup}$ object \citep{chabrier2000a}, with a minimum rotational period of 8 hours, the spin velocity is about $19.6-21\rm\,kms^{-1}$. 
A longer period (P $>8$ hours) will result in a slower spin.
At the age of ABDMG, which is $\sim120-200$ Myrs, our upper limit of spin rate of GU\,Psc\,b is similar with that of other planetary-mass companions at different ages (3-300 Myrs old) \citep[e.g.,][]{snellen2014,zhou2016,biller2018,bryan2018}.
Because the radii contract along with the loss of interior entropy for young objects, the spin velocities increase as the objects cool with age under the conservation of angular momentum.
Therefore, the spin rates of GU Psc b could reach as high as $30\rm\,kms^{-1}$ after radius contracting to one Jupiter radius. 
A better period constraint is required to test if the spin rate is consistent with the suggested universal spin-mass relation of \citet[][]{scholz2018}(see also \citealt{zhou2019}). 

\subsection{Gray modulations and atmospheric heterogeneity}
Both the color-magnitude variations shown in Figure~\ref{fig:cmd} and the ratio of the brightest--to--dimmest spectra shown in Figure~\ref{fig:spec} demonstrate that the modulation amplitudes of GU\,Psc\,b are similar in $J'$ and $H'$ band (peak-to-trough variation of $2.6 \pm 0.9\%$ and $3.2  \pm 0.9\%$, respectively), and hence mostly gray.
Previous observations find similar modulation amplitudes in $J$ and $H$ bands for L/T transition objects \citep[e.g.,][]{artigau2009,radigan2012,apai2013}, although the modulation amplitudes can be different in molecular bands (e.g., water) or at longer wavelengths (e.g., $\rm K_{s}$ band) \citep[e.g.,][]{apai2013,biller2013}. 
The modeling of these data suggest that spatial variations of cloud properties are responsible for the modulations \citep[e.g.,][]{apai2013}. Detailed atmospheric modeling of space-based, high-precision time-domain data argues for {\em simultaneous} changes in cloud thickness {\em and} temperature as the cause for spatially varying cloud brightness (thin warm and thick cold clouds, \citealt{apai2013,buenzli2015a}). 
GU\,Psc\,b also shows similarly gray rotational modulations. Therefore, we argue that correlated variations of cloud opacity and temperature are the most likely cause for the observed gray modulations in GU\,Psc\,b. 

The observed weak (non-zero) wavelength dependence may also carry information about spatial variation in particle size distribution.
A weak and positive wavelength dependence in rotational modulations could be explained by the varying presence of particles with sizes larger than one micron \citep[e.g. see Figure 5][]{hiranaka2016}. If this is true, then GU\,Psc\,b's atmosphere is in contrast to L dwarf atmospheres, which are often found to be reddened or extinguished by sub-micron grains \citep[][see also retrieval analysis from \citealt{burningham2017}]{hiranaka2016,marocco2014}.
This would be consistent with the cloud thinning scenario that predicts larger mean particle sizes (or larger $f_{\rm sed}$)  for mid-T and smaller mean particle sizes (or smaller $f_{\rm sed}$) for L dwarfs \citep{saumon2008}.

\subsection{No strong color change in rotational modulations across the L/T Transition}
In the color-magnitude plot shown in Figure~\ref{fig:cmd} even the most highly variable object (2M2139) shows only less than 20\% relative difference between its $J$-band and the $H$-band modulation amplitudes ($31.4 \pm 0.1\% $ vs. $ 27.6 \pm 0.1\%$). All other objects show similarly gray color change in modulations ($\Delta (J'-H') \ll \Delta J'$): no known source is varying its color at a level comparable to the brightness modulations.
In other words, the observed modulations in all sources are close to gray, albeit with some variety.
The lack of strong color in rotational modulations in brown dwarfs with spectral types ranging from mid-L to late-T is consistent with the paradigm in which spatially heterogeneous clouds modulate the hemisphere-integrated brightness.

Given that the color changes due to atmospheric heterogeneity are different from the overall blue--to--red color evolution found across the L/T transition (a large color evolution with only a small change in absolute $J$-band magnitude from L8-to-T5 spectral type), we conclude that atmospheric heterogeneity alone does not directly cause the drastic color evolution across the L/T transition, at least for low-gravity atmospheres (i.e., GU\,Psc\,b, SIMP0136, PSOJ318).
However, as previously suggested by \citet{radigan2012}, atmospheric heterogeneity could still affect the evolution of atmospheres {\it over evolutionary timescales} --- an atmosphere with thinner clouds or more patchy cloud distribution cools more efficiently because more flux can be radiated from below the cloud base.
More efficient cooling leads to a larger loss of entropy over evolutionary timescales.
As a result, the loss of interior entropy is coupled with varying degrees of atmospheric heterogeneity and could still lead to the observed drastic color evolution in L/T transition.
Our observations only probe the impact of atmospheric heterogeneity with fixed interior entropy in rotational timescales.
Change of cloud structure will affect the relative abundance of objects with different spectral sub-types across the L/T transition, as predicted by \citet{saumon2008}. Observational-bias corrected samples of brown dwarfs will be powerful to test the coupled evolution of cloud structure and interior entropy over evolutionary timescales.

It is tempting to perceive a possible trend in Figure~\ref{fig:cmd} in the slope of the magnitude-color variations from L to T dwarfs: L dwarfs become brighter and bluer, early T dwarfs become brighter with almost no color change, and the two out of three mid-to-late T dwarfs become brighter and redder. However, we are limited by the small sample size of T dwarfs with time-resolved spectrophotometry, and the uncertain slope of GU\,Psc\,b due to incomplete phase coverage. 
Therefore, we remain cautious about the significance of this tentative trend. More long-term time-resolved spectroscopy of T dwarfs is needed to verify this tentative trend and to test if there is a statistically significant difference between L and T dwarfs in the nature of their wavelength-dependent rotational modulations.

\section{Conclusions}\label{s:Conclusions}
In this study we present the HST time-resolved near-infrared spectral variations of the planetary-mass, T3.5 spectral type object GU\,Psc\,b. The key conclusions of our study are as follows:

\begin{enumerate}
    \item We confirm the previously reported \citep{naud2017} tentative rotational modulations in the planetary-mass companion GU\,Psc\,b. This is the first planetary-mass object in the T3--T5 spectral range with confirmed rotational modulations.
  
    \item Based on our HST WFC3/G141 observations we place a lower limit of $2.7 \pm 0.8\%$ peak-to-trough flux variation of GU\,Psc\,b and a period of eight hours or longer. As our phase coverage is incomplete, it is likely that the actual flux variations are somewhat higher.
  
    \item We find mostly gray (wavelength-dependent slope of m = $0.025 \pm 0.020\rm \,\mu m^{-1}$, c.f. Figure \ref{fig:spec}) rotational modulations for wavelengths from $1.1$ to $1.67\rm \,\mu m$ excluding the water-band. Based on the gray modulations, we argue that cloud opacity likely dominates the rotational modulations in the photosphere of low-gravity mid-T dwarf GU\,Psc\,b.

    \item From our compilation of mid-L to late-T dwarfs, we find their rotational modulations to be mostly gray, including objects across the L/T transition. We argue that atmospheric heterogeneity cannot explain the drastic color evolution across the L/T transition over rotational timescales. Cloud heterogeneity could still play an important role in atmospheric evolution in the L/T transition over evolutionary timescales. 
    
    \item From L to T dwarfs we find an interesting but tentative trend in the slope of the magnitude-color variations.  If confirmed, this trend would indicate that the nature of the rotational modulation is spectral type-dependent. However, more samples of T dwarf's rotational modulations with complete phase coverage are needed to test the significance of the trend.
\end{enumerate}
 Together with \object{2M1207b}, 	
\object{2MASS J13243553+6358281}, PSOJ318, Ross 458c, and SIMP0136, GU\,Psc\,b is another rare planetary-mass objects with large infrared modulation amplitude. This object is a yet unique example of T3-T5 spectral type, low-gravity object with detected rotational modulations.  As such it provides an important reference to study cloud-structure evolution as functions of effective temperature and gravity.
 Soon the higher sensitivity and wider wavelength coverage of next-generation telescopes such as the {\it  James Webb Space Telescope} (JWST) and {\it Extremely Large Telescopes} (Giant Magellan Telescope, Thirty Meter Telescope, and the European ELT) will transform time-resolved spectroscopy into an even more powerful method for constraining the cloud structure and particle-size distribution.
\section{Acknowledgement}
We would like to thank the anonymous referee for constructive comments and suggestions that significantly improve this paper. We would like to thank Esther Buenzli for providing the time-resolved spectra of 2M2228.
Support for Program number HST-GO-14241.001A was provided by NASA through a grant from the Space Telescope Science Institute, which is operated by the Association of Universities for Research in Astronomy, Incorporated, under NASA contract NAS5-26555.
This research has benefited from the Montreal Brown Dwarf and Exoplanet Spectral Library, maintained by Jonathan Gagn\'e, and the SpeX Prism Spectral Libraries, maintained by Adam Burgasser at \url{http://pono.ucsd.edu/~adam/browndwarfs/spexprism.}
This publication makes use of data products from the Two Micron All Sky Survey, which is a joint project of the University of Massachusetts and the Infrared Processing and Analysis Center/California Institute of Technology, funded by the National Aeronautics and Space Administration and the National Science Foundation.
This work has made use of data from the European Space Agency (ESA) mission
{\it Gaia} (\url{https://www.cosmos.esa.int/gaia}), processed by the {\it Gaia}
Data Processing and Analysis Consortium (DPAC,
\url{https://www.cosmos.esa.int/web/gaia/dpac/consortium}). Funding for the DPAC
has been provided by national institutions, in particular the institutions
participating in the {\it Gaia} Multilateral Agreement.

\bibliography{browndwarf.bib}

\begin{thebibliography}{}
\expandafter\ifx\csname natexlab\endcsname\relax\def\natexlab#1{#1}\fi
\providecommand{\url}[1]{\href{#1}{#1}}

\bibitem[{{Ackerman} \& {Marley}(2001)}]{ackerman2001}
{Ackerman}, A.~S., \& {Marley}, M.~S. 2001, \apj, 556, 872

\bibitem[{{Apai} {et~al.}(2013){Apai}, {Radigan}, {Buenzli}, {Burrows}, {Reid},
  \& {Jayawardhana}}]{apai2013}
{Apai}, D., {Radigan}, J., {Buenzli}, E., {et~al.} 2013, \apj, 768, 121

\bibitem[{{Apai} {et~al.}(2017){Apai}, {Karalidi}, {Marley}, {Yang}, {Flateau},
  {Metchev}, {Cowan}, {Buenzli}, {Burgasser}, {Radigan}, {Artigau}, \&
  {Lowrance}}]{apai2017}
{Apai}, D., {Karalidi}, T., {Marley}, M.~S., {et~al.} 2017, Science, 357, 683

\bibitem[{{Artigau}(2018)}]{artigau2018}
{Artigau}, {\'E}. 2018, ArXiv e-prints, arXiv:1803.07672

\bibitem[{{Artigau} {et~al.}(2009){Artigau}, {Bouchard}, {Doyon}, \&
  {Lafreni{\`e}re}}]{artigau2009}
{Artigau}, {\'E}., {Bouchard}, S., {Doyon}, R., \& {Lafreni{\`e}re}, D. 2009,
  \apj, 701, 1534

\bibitem[{{Baraffe} {et~al.}(2003){Baraffe}, {Chabrier}, {Barman}, {Allard}, \&
  {Hauschildt}}]{baraffe2003}
{Baraffe}, I., {Chabrier}, G., {Barman}, T.~S., {Allard}, F., \& {Hauschildt},
  P.~H. 2003, \aap, 402, 701

\bibitem[{{Barrado y Navascu{\'e}s} \& {Mart{\'\i}n}(2003)}]{barrado2003}
{Barrado y Navascu{\'e}s}, D., \& {Mart{\'\i}n}, E.~L. 2003, \aj, 126, 2997

\bibitem[{{Berta} {et~al.}(2012){Berta}, {Charbonneau}, {D{\'e}sert},
  {Miller-Ricci Kempton}, {McCullough}, {Burke}, {Fortney}, {Irwin}, {Nutzman},
  \& {Homeier}}]{berta2012}
{Berta}, Z.~K., {Charbonneau}, D., {D{\'e}sert}, J.-M., {et~al.} 2012, \apj,
  747, 35

\bibitem[{{Bertin} \& {Arnouts}(1996)}]{bertin1996}
{Bertin}, E., \& {Arnouts}, S. 1996, \aaps, 117, 393

\bibitem[{{Best} {et~al.}(2015){Best}, {Liu}, {Magnier}, {Deacon}, {Aller},
  {Redstone}, {Burgett}, {Chambers}, {Draper}, {Flewelling}, {Hodapp},
  {Kaiser}, {Metcalfe}, {Tonry}, {Wainscoat}, \& {Waters}}]{best2015}
{Best}, W. M.~J., {Liu}, M.~C., {Magnier}, E.~A., {et~al.} 2015, \apj, 814, 118

\bibitem[{{Biller}(2017)}]{biller2017}
{Biller}, B. 2017, The Astronomical Review, 13, 1

\bibitem[{{Biller} {et~al.}(2013){Biller}, {Crossfield}, {Mancini}, {Ciceri},
  {Southworth}, {Kopytova}, {Bonnefoy}, {Deacon}, {Schlieder}, {Buenzli},
  {Brandner}, {Allard}, {Homeier}, {Freytag}, {Bailer-Jones}, {Greiner},
  {Henning}, \& {Goldman}}]{biller2013}
{Biller}, B.~A., {Crossfield}, I.~J.~M., {Mancini}, L., {et~al.} 2013, \apjl,
  778, L10

\bibitem[{{Biller} {et~al.}(2015){Biller}, {Vos}, {Bonavita}, {Buenzli},
  {Baxter}, {Crossfield}, {Allers}, {Liu}, {Bonnefoy}, {Deacon}, {Brandner},
  {Schlieder}, {Dupuy}, {Kopytova}, {Manjavacas}, {Allard}, {Homeier}, \&
  {Henning}}]{biller2015}
{Biller}, B.~A., {Vos}, J., {Bonavita}, M., {et~al.} 2015, \apjl, 813, L23

\bibitem[{{Biller} {et~al.}(2018){Biller}, {Vos}, {Buenzli}, {Allers},
  {Bonnefoy}, {Charnay}, {B{\'e}zard}, {Allard}, {Homeier}, {Bonavita},
  {Brandner}, {Crossfield}, {Dupuy}, {Henning}, {Kopytova}, {Liu},
  {Manjavacas}, \& {Schlieder}}]{biller2018}
{Biller}, B.~A., {Vos}, J., {Buenzli}, E., {et~al.} 2018, \aj, 155, 95

\bibitem[{{Bowler} {et~al.}(2013){Bowler}, {Liu}, {Shkolnik}, \&
  {Dupuy}}]{bowler2013}
{Bowler}, B.~P., {Liu}, M.~C., {Shkolnik}, E.~L., \& {Dupuy}, T.~J. 2013, \apj,
  774, 55

\bibitem[{{Bryan} {et~al.}(2018){Bryan}, {Benneke}, {Knutson}, {Batygin}, \&
  {Bowler}}]{bryan2018}
{Bryan}, M.~L., {Benneke}, B., {Knutson}, H.~A., {Batygin}, K., \& {Bowler},
  B.~P. 2018, Nature Astronomy, 2, 138

\bibitem[{{Buenzli} {et~al.}(2014){Buenzli}, {Apai}, {Radigan}, {Reid}, \&
  {Flateau}}]{buenzli2014}
{Buenzli}, E., {Apai}, D., {Radigan}, J., {Reid}, I.~N., \& {Flateau}, D. 2014,
  \apj, 782, 77

\bibitem[{{Buenzli} {et~al.}(2015){Buenzli}, {Saumon}, {Marley}, {Apai},
  {Radigan}, {Bedin}, {Reid}, \& {Morley}}]{buenzli2015a}
{Buenzli}, E., {Saumon}, D., {Marley}, M.~S., {et~al.} 2015, \apj, 798, 127

\bibitem[{{Buenzli} {et~al.}(2012){Buenzli}, {Apai}, {Morley}, {Flateau},
  {Showman}, {Burrows}, {Marley}, {Lewis}, \& {Reid}}]{buenzli2012}
{Buenzli}, E., {Apai}, D., {Morley}, C.~V., {et~al.} 2012, \apjl, 760, L31

\bibitem[{{Burgasser}(2014)}]{burgasser2014a}
{Burgasser}, A.~J. 2014, in Astronomical Society of India Conference Series,
  Vol.~11, Astronomical Society of India Conference Series, 7--16

\bibitem[{{Burgasser} {et~al.}(2006{\natexlab{a}}){Burgasser}, {Geballe},
  {Leggett}, {Kirkpatrick}, \& {Golimowski}}]{burgasser2006a}
{Burgasser}, A.~J., {Geballe}, T.~R., {Leggett}, S.~K., {Kirkpatrick}, J.~D.,
  \& {Golimowski}, D.~A. 2006{\natexlab{a}}, \apj, 637, 1067

\bibitem[{{Burgasser} {et~al.}(2006{\natexlab{b}}){Burgasser}, {Kirkpatrick},
  {Cruz}, {Reid}, {Leggett}, {Liebert}, {Burrows}, \& {Brown}}]{burgasser2006b}
{Burgasser}, A.~J., {Kirkpatrick}, J.~D., {Cruz}, K.~L., {et~al.}
  2006{\natexlab{b}}, \apjs, 166, 585

\bibitem[{{Burgasser} {et~al.}(2002){Burgasser}, {Marley}, {Ackerman},
  {Saumon}, {Lodders}, {Dahn}, {Harris}, \& {Kirkpatrick}}]{burgasser2002}
{Burgasser}, A.~J., {Marley}, M.~S., {Ackerman}, A.~S., {et~al.} 2002, \apjl,
  571, L151

\bibitem[{{Burgasser} {et~al.}(2004){Burgasser}, {McElwain}, {Kirkpatrick},
  {Cruz}, {Tinney}, \& {Reid}}]{burgasser2004}
{Burgasser}, A.~J., {McElwain}, M.~W., {Kirkpatrick}, J.~D., {et~al.} 2004,
  \aj, 127, 2856

\bibitem[{{Burningham} {et~al.}(2017){Burningham}, {Marley}, {Line}, {Lupu},
  {Visscher}, {Morley}, {Saumon}, \& {Freedman}}]{burningham2017}
{Burningham}, B., {Marley}, M.~S., {Line}, M.~R., {et~al.} 2017, \mnras, 470,
  1177

\bibitem[{{Burrows} {et~al.}(2006){Burrows}, {Sudarsky}, \&
  {Hubeny}}]{burrows2006}
{Burrows}, A., {Sudarsky}, D., \& {Hubeny}, I. 2006, \apj, 640, 1063

\bibitem[{{Chabrier} {et~al.}(2000){Chabrier}, {Baraffe}, {Allard}, \&
  {Hauschildt}}]{chabrier2000a}
{Chabrier}, G., {Baraffe}, I., {Allard}, F., \& {Hauschildt}, P. 2000, \apj,
  542, 464

\bibitem[{{Charnay} {et~al.}(2018){Charnay}, {B{\'e}zard}, {Baudino},
  {Bonnefoy}, {Boccaletti}, \& {Galicher}}]{charnay2018}
{Charnay}, B., {B{\'e}zard}, B., {Baudino}, J.-L., {et~al.} 2018, \apj, 854,
  172

\bibitem[{{Cowan} \& {Agol}(2008)}]{cowan2008}
{Cowan}, N.~B., \& {Agol}, E. 2008, \apjl, 678, L129

\bibitem[{{Cowan} {et~al.}(2013){Cowan}, {Fuentes}, \& {Haggard}}]{cowan2013}
{Cowan}, N.~B., {Fuentes}, P.~A., \& {Haggard}, H.~M. 2013, \mnras, 434, 2465

\bibitem[{{Cutri} {et~al.}(2003){Cutri}, {Skrutskie}, {van Dyk}, {Beichman},
  {Carpenter}, {Chester}, {Cambresy}, {Evans}, {Fowler}, {Gizis}, {Howard},
  {Huchra}, {Jarrett}, {Kopan}, {Kirkpatrick}, {Light}, {Marsh}, {McCallon},
  {Schneider}, {Stiening}, {Sykes}, {Weinberg}, {Wheaton}, {Wheelock}, \&
  {Zacarias}}]{cutri2003}
{Cutri}, R.~M., {Skrutskie}, M.~F., {van Dyk}, S., {et~al.} 2003, VizieR Online
  Data Catalog, II/246

\bibitem[{{Dupuy} \& {Liu}(2012)}]{dupuy2012}
{Dupuy}, T.~J., \& {Liu}, M.~C. 2012, \apjs, 201, 19

\bibitem[{{Dupuy} {et~al.}(2009){Dupuy}, {Liu}, \& {Ireland}}]{dupuy2009}
{Dupuy}, T.~J., {Liu}, M.~C., \& {Ireland}, M.~J. 2009, \apj, 699, 168

\bibitem[{{Gaia Collaboration} {et~al.}(2016){Gaia Collaboration}, {Prusti},
  {de Bruijne}, {Brown}, {Vallenari}, {Babusiaux}, {Bailer-Jones}, {Bastian},
  {Biermann}, {Evans}, {Eyer}, {Jansen}, {Jordi}, {Klioner}, {Lammers},
  {Lindegren}, {Luri}, {Mignard}, {Milligan}, {Panem}, {Poinsignon},
  {Pourbaix}, {Randich}, {Sarri}, {Sartoretti}, {Siddiqui}, {Soubiran},
  {Valette}, {van Leeuwen}, {Walton}, {Aerts}, {Arenou}, {Cropper}, {Drimmel},
  {H{\o}g}, {Katz}, {Lattanzi}, {O'Mullane}, {Grebel}, {Holland}, {Huc},
  {Passot}, {Bramante}, {Cacciari}, {Casta{\~n}eda}, {Chaoul}, {Cheek}, {De
  Angeli}, {Fabricius}, {Guerra}, {Hern{\'a}ndez}, {Jean-Antoine-Piccolo},
  {Masana}, {Messineo}, {Mowlavi}, {Nienartowicz}, {Ord{\'o}{\~n}ez-Blanco},
  {Panuzzo}, {Portell}, {Richards}, {Riello}, {Seabroke}, {Tanga},
  {Th{\'e}venin}, {Torra}, {Els}, {Gracia-Abril}, {Comoretto},
  {Garcia-Reinaldos}, {Lock}, {Mercier}, {Altmann}, {Andrae}, {Astraatmadja},
  {Bellas-Velidis}, {Benson}, {Berthier}, {Blomme}, {Busso}, {Carry},
  {Cellino}, {Clementini}, {Cowell}, {Creevey}, {Cuypers}, {Davidson}, {De
  Ridder}, {de Torres}, {Delchambre}, {Dell'Oro}, {Ducourant}, {Fr{\'e}mat},
  {Garc{\'\i}a-Torres}, {Gosset}, {Halbwachs}, {Hambly}, {Harrison}, {Hauser},
  {Hestroffer}, {Hodgkin}, {Huckle}, {Hutton}, {Jasniewicz}, {Jordan},
  {Kontizas}, {Korn}, {Lanzafame}, {Manteiga}, {Moitinho}, {Muinonen},
  {Osinde}, {Pancino}, {Pauwels}, {Petit}, {Recio-Blanco}, {Robin}, {Sarro},
  {Siopis}, {Smith}, {Smith}, {Sozzetti}, {Thuillot}, {van Reeven}, {Viala},
  {Abbas}, {Abreu Aramburu}, {Accart}, {Aguado}, {Allan}, {Allasia},
  {Altavilla}, {{\'A}lvarez}, {Alves}, {Anderson}, {Andrei}, {Anglada Varela},
  {Antiche}, {Antoja}, {Ant{\'o}n}, {Arcay}, {Atzei}, {Ayache}, {Bach},
  {Baker}, {Balaguer-N{\'u}{\~n}ez}, {Barache}, {Barata}, {Barbier}, {Barblan},
  {Baroni}, {Barrado y Navascu{\'e}s}, {Barros}, {Barstow}, {Becciani},
  {Bellazzini}, {Bellei}, {Bello Garc{\'\i}a}, {Belokurov}, {Bendjoya},
  {Berihuete}, {Bianchi}, {Bienaym{\'e}}, {Billebaud}, {Blagorodnova},
  {Blanco-Cuaresma}, {Boch}, {Bombrun}, {Borrachero}, {Bouquillon}, {Bourda},
  {Bouy}, {Bragaglia}, {Breddels}, {Brouillet}, {Br{\"u}semeister},
  {Bucciarelli}, {Budnik}, {Burgess}, {Burgon}, {Burlacu}, {Busonero}, {Buzzi},
  {Caffau}, {Cambras}, {Campbell}, {Cancelliere}, {Cantat-Gaudin}, {Carlucci},
  {Carrasco}, {Castellani}, {Charlot}, {Charnas}, {Charvet}, {Chassat},
  {Chiavassa}, {Clotet}, {Cocozza}, {Collins}, {Collins}, {Costigan}, {Crifo},
  {Cross}, {Crosta}, {Crowley}, {Dafonte}, {Damerdji}, {Dapergolas}, {David},
  {David}, {De Cat}, {de Felice}, {de Laverny}, {De Luise}, {De March}, {de
  Martino}, {de Souza}, {Debosscher}, {del Pozo}, {Delbo}, {Delgado},
  {Delgado}, {di Marco}, {Di Matteo}, {Diakite}, {Distefano}, {Dolding}, {Dos
  Anjos}, {Drazinos}, {Dur{\'a}n}, {Dzigan}, {Ecale}, {Edvardsson}, {Enke},
  {Erdmann}, {Escolar}, {Espina}, {Evans}, {Eynard Bontemps}, {Fabre},
  {Fabrizio}, {Faigler}, {Falc{\~a}o}, {Farr{\`a}s Casas}, {Faye}, {Federici},
  {Fedorets}, {Fern{\'a}ndez-Hern{\'a}ndez}, {Fernique}, {Fienga}, {Figueras},
  {Filippi}, {Findeisen}, {Fonti}, {Fouesneau}, {Fraile}, {Fraser}, {Fuchs},
  {Furnell}, {Gai}, {Galleti}, {Galluccio}, {Garabato}, {Garc{\'\i}a-Sedano},
  {Gar{\'e}}, {Garofalo}, {Garralda}, {Gavras}, {Gerssen}, {Geyer}, {Gilmore},
  {Girona}, {Giuffrida}, {Gomes}, {Gonz{\'a}lez-Marcos},
  {Gonz{\'a}lez-N{\'u}{\~n}ez}, {Gonz{\'a}lez-Vidal}, {Granvik}, {Guerrier},
  {Guillout}, {Guiraud}, {G{\'u}rpide}, {Guti{\'e}rrez-S{\'a}nchez}, {Guy},
  {Haigron}, {Hatzidimitriou}, {Haywood}, {Heiter}, {Helmi}, {Hobbs},
  {Hofmann}, {Holl}, {Holland }, {Hunt}, {Hypki}, {Icardi}, {Irwin}, {Jevardat
  de Fombelle}, {Jofr{\'e}}, {Jonker}, {Jorissen}, {Julbe}, {Karampelas},
  {Kochoska}, {Kohley}, {Kolenberg}, {Kontizas}, {Koposov}, {Kordopatis},
  {Koubsky}, {Kowalczyk}, {Krone-Martins}, {Kudryashova}, {Kull}, {Bachchan},
  {Lacoste-Seris}, {Lanza}, {Lavigne}, {Le Poncin-Lafitte}, {Lebreton},
  {Lebzelter}, {Leccia}, {Leclerc}, {Lecoeur-Taibi}, {Lemaitre}, {Lenhardt},
  {Leroux}, {Liao}, {Licata}, {Lindstr{\o}m}, {Lister}, {Livanou}, {Lobel},
  {L{\"o}ffler}, {L{\'o}pez}, {Lopez-Lozano}, {Lorenz}, {Loureiro},
  {MacDonald}, {Magalh{\~a}es Fernandes}, {Managau}, {Mann}, {Mantelet},
  {Marchal}, {Marchant}, {Marconi}, {Marie}, {Marinoni}, {Marrese},
  {Marschalk{\'o}}, {Marshall}, {Mart{\'\i}n-Fleitas}, {Martino}, {Mary},
  {Matijevi{\v{c}}}, {Mazeh}, {McMillan}, {Messina}, {Mestre}, {Michalik},
  {Millar}, {Miranda}, {Molina}, {Molinaro}, {Molinaro}, {Moln{\'a}r},
  {Moniez}, {Montegriffo}, {Monteiro}, {Mor}, {Mora}, {Morbidelli}, {Morel},
  {Morgenthaler}, {Morley}, {Morris}, {Mulone}, {Muraveva}, {Musella},
  {Narbonne}, {Nelemans}, {Nicastro}, {Noval}, {Ord{\'e}novic},
  {Ordieres-Mer{\'e}}, {Osborne}, {Pagani}, {Pagano}, {Pailler}, {Palacin},
  {Palaversa}, {Parsons}, {Paulsen}, {Pecoraro}, {Pedrosa}, {Pentik{\"a}inen},
  {Pereira}, {Pichon}, {Piersimoni}, {Pineau}, {Plachy}, {Plum}, {Poujoulet},
  {Pr{\v{s}}a}, {Pulone}, {Ragaini}, {Rago}, {Rambaux}, {Ramos-Lerate},
  {Ranalli}, {Rauw}, {Read}, {Regibo}, {Renk}, {Reyl{\'e}}, {Ribeiro},
  {Rimoldini}, {Ripepi}, {Riva}, {Rixon}, {Roelens}, {Romero-G{\'o}mez},
  {Rowell}, {Royer}, {Rudolph}, {Ruiz-Dern}, {Sadowski}, {Sagrist{\`a}
  Sell{\'e}s}, {Sahlmann}, {Salgado}, {Salguero}, {Sarasso}, {Savietto},
  {Schnorhk}, {Schultheis}, {Sciacca}, {Segol}, {Segovia}, {Segransan},
  {Serpell}, {Shih}, {Smareglia}, {Smart}, {Smith}, {Solano}, {Solitro},
  {Sordo}, {Soria Nieto}, {Souchay}, {Spagna}, {Spoto}, {Stampa}, {Steele},
  {Steidelm{\"u}ller}, {Stephenson}, {Stoev}, {Suess}, {S{\"u}veges}, {Surdej},
  {Szabados}, {Szegedi-Elek}, {Tapiador}, {Taris}, {Tauran}, {Taylor},
  {Teixeira}, {Terrett}, {Tingley}, {Trager}, {Turon}, {Ulla}, {Utrilla},
  {Valentini}, {van Elteren}, {Van Hemelryck}, {van Leeuwen}, {Varadi},
  {Vecchiato}, {Veljanoski}, {Via}, {Vicente}, {Vogt}, {Voss}, {Votruba},
  {Voutsinas}, {Walmsley}, {Weiler}, {Weingrill}, {Werner}, {Wevers},
  {Whitehead}, {Wyrzykowski}, {Yoldas}, {{\v{Z}}erjal}, {Zucker}, {Zurbach},
  {Zwitter}, {Alecu}, {Allen}, {Allende Prieto}, {Amorim},
  {Anglada-Escud{\'e}}, {Arsenijevic}, {Azaz}, {Balm}, {Beck}, {Bernstein},
  {Bigot}, {Bijaoui}, {Blasco}, {Bonfigli}, {Bono}, {Boudreault}, {Bressan},
  {Brown}, {Brunet}, {Bunclark}, {Buonanno}, {Butkevich}, {Carret}, {Carrion},
  {Chemin}, {Ch{\'e}reau}, {Corcione}, {Darmigny}, {de Boer}, {de Teodoro}, {de
  Zeeuw}, {Delle Luche}, {Domingues}, {Dubath}, {Fodor}, {Fr{\'e}zouls},
  {Fries}, {Fustes}, {Fyfe}, {Gallardo}, {Gallegos}, {Gardiol}, {Gebran},
  {Gomboc}, {G{\'o}mez}, {Grux}, {Gueguen}, {Heyrovsky}, {Hoar}, {Iannicola},
  {Isasi Parache}, {Janotto}, {Joliet}, {Jonckheere}, {Keil}, {Kim},
  {Klagyivik}, {Klar}, {Knude}, {Kochukhov}, {Kolka}, {Kos}, {Kutka}, {Lainey},
  {LeBouquin}, {Liu}, {Loreggia}, {Makarov}, {Marseille}, {Martayan},
  {Martinez-Rubi}, {Massart}, {Meynadier}, {Mignot}, {Munari}, {Nguyen},
  {Nordlander}, {Ocvirk}, {O'Flaherty}, {Olias Sanz}, {Ortiz}, {Osorio},
  {Oszkiewicz}, {Ouzounis}, {Palmer}, {Park}, {Pasquato}, {Peltzer}, {Peralta},
  {P{\'e}turaud}, {Pieniluoma}, {Pigozzi}, {Poels}, {Prat}, {Prod'homme},
  {Raison}, {Rebordao}, {Risquez}, {Rocca-Volmerange}, {Rosen}, {Ruiz-Fuertes},
  {Russo}, {Sembay}, {Serraller Vizcaino}, {Short}, {Siebert}, {Silva},
  {Sinachopoulos}, {Slezak}, {Soffel}, {Sosnowska}, {Strai{\v{z}}ys}, {ter
  Linden}, {Terrell}, {Theil}, {Tiede}, {Troisi}, {Tsalmantza}, {Tur},
  {Vaccari}, {Vachier}, {Valles}, {Van Hamme}, {Veltz}, {Virtanen}, {Wallut},
  {Wichmann}, {Wilkinson}, {Ziaeepour}, \& {Zschocke}}]{GaiaCollaboration2016}
{Gaia Collaboration}, {Prusti}, T., {de Bruijne}, J.~H.~J., {et~al.} 2016,
  \aap, 595, A1

\bibitem[{{Gaia Collaboration} {et~al.}(2018){Gaia Collaboration}, {Brown},
  {Vallenari}, {Prusti}, {de Bruijne}, {Babusiaux}, {Bailer-Jones}, {Biermann},
  {Evans}, {Eyer}, {Jansen}, {Jordi}, {Klioner}, {Lammers}, {Lindegren},
  {Luri}, {Mignard}, {Panem}, {Pourbaix}, {Randich}, {Sartoretti}, {Siddiqui},
  {Soubiran}, {van Leeuwen}, {Walton}, {Arenou}, {Bastian}, {Cropper},
  {Drimmel}, {Katz}, {Lattanzi}, {Bakker}, {Cacciari}, {Casta{\~n}eda},
  {Chaoul}, {Cheek}, {De Angeli}, {Fabricius}, {Guerra}, {Holl}, {Masana},
  {Messineo}, {Mowlavi}, {Nienartowicz}, {Panuzzo}, {Portell}, {Riello},
  {Seabroke}, {Tanga}, {Th{\'e}venin}, {Gracia-Abril}, {Comoretto},
  {Garcia-Reinaldos}, {Teyssier}, {Altmann}, {Andrae}, {Audard},
  {Bellas-Velidis}, {Benson}, {Berthier}, {Blomme}, {Burgess}, {Busso},
  {Carry}, {Cellino}, {Clementini}, {Clotet}, {Creevey}, {Davidson}, {De
  Ridder}, {Delchambre}, {Dell'Oro}, {Ducourant},
  {Fern{\'a}ndez-Hern{\'a}ndez}, {Fouesneau}, {Fr{\'e}mat}, {Galluccio},
  {Garc{\'\i}a-Torres}, {Gonz{\'a}lez-N{\'u}{\~n}ez}, {Gonz{\'a}lez-Vidal},
  {Gosset}, {Guy}, {Halbwachs}, {Hambly}, {Harrison}, {Hern{\'a}ndez},
  {Hestroffer}, {Hodgkin}, {Hutton}, {Jasniewicz}, {Jean-Antoine-Piccolo},
  {Jordan}, {Korn}, {Krone-Martins}, {Lanzafame}, {Lebzelter}, {L{\"o}ffler},
  {Manteiga}, {Marrese}, {Mart{\'\i}n-Fleitas}, {Moitinho}, {Mora}, {Muinonen},
  {Osinde}, {Pancino}, {Pauwels}, {Petit}, {Recio-Blanco}, {Richards},
  {Rimoldini}, {Robin}, {Sarro}, {Siopis}, {Smith}, {Sozzetti}, {S{\"u}veges},
  {Torra}, {van Reeven}, {Abbas}, {Abreu Aramburu}, {Accart}, {Aerts},
  {Altavilla}, {{\'A}lvarez}, {Alvarez}, {Alves}, {Anderson}, {Andrei},
  {Anglada Varela}, {Antiche}, {Antoja}, {Arcay}, {Astraatmadja}, {Bach},
  {Baker}, {Balaguer-N{\'u}{\~n}ez}, {Balm}, {Barache}, {Barata}, {Barbato},
  {Barblan}, {Barklem}, {Barrado}, {Barros}, {Barstow}, {Bartholom{\'e}
  Mu{\~n}oz}, {Bassilana}, {Becciani}, {Bellazzini}, {Berihuete}, {Bertone},
  {Bianchi}, {Bienaym{\'e}}, {Blanco-Cuaresma}, {Boch}, {Boeche}, {Bombrun},
  {Borrachero}, {Bossini}, {Bouquillon}, {Bourda}, {Bragaglia}, {Bramante},
  {Breddels}, {Bressan}, {Brouillet}, {Br{\"u}semeister}, {Brugaletta},
  {Bucciarelli}, {Burlacu}, {Busonero}, {Butkevich}, {Buzzi}, {Caffau},
  {Cancelliere}, {Cannizzaro}, {Cantat-Gaudin}, {Carballo}, {Carlucci},
  {Carrasco}, {Casamiquela}, {Castellani}, {Castro-Ginard}, {Charlot},
  {Chemin}, {Chiavassa}, {Cocozza}, {Costigan}, {Cowell}, {Crifo}, {Crosta},
  {Crowley}, {Cuypers}, {Dafonte}, {Damerdji}, {Dapergolas}, {David}, {David},
  {de Laverny}, {De Luise}, {De March}, {de Martino}, {de Souza}, {de Torres},
  {Debosscher}, {del Pozo}, {Delbo}, {Delgado}, {Delgado}, {Di Matteo},
  {Diakite}, {Diener}, {Distefano}, {Dolding}, {Drazinos}, {Dur{\'a}n},
  {Edvardsson}, {Enke}, {Eriksson}, {Esquej}, {Eynard Bontemps}, {Fabre},
  {Fabrizio}, {Faigler}, {Falc{\~a}o}, {Farr{\`a}s Casas}, {Federici},
  {Fedorets}, {Fernique}, {Figueras}, {Filippi}, {Findeisen}, {Fonti},
  {Fraile}, {Fraser}, {Fr{\'e}zouls}, {Gai}, {Galleti}, {Garabato},
  {Garc{\'\i}a-Sedano}, {Garofalo}, {Garralda}, {Gavel}, {Gavras}, {Gerssen},
  {Geyer}, {Giacobbe}, {Gilmore}, {Girona}, {Giuffrida}, {Glass}, {Gomes},
  {Granvik}, {Gueguen}, {Guerrier}, {Guiraud}, {Guti{\'e}rrez-S{\'a}nchez},
  {Haigron}, {Hatzidimitriou}, {Hauser}, {Haywood}, {Heiter}, {Helmi}, {Heu},
  {Hilger}, {Hobbs}, {Hofmann}, {Holland}, {Huckle}, {Hypki}, {Icardi},
  {Jan{\ss}en}, {Jevardat de Fombelle}, {Jonker}, {Juh{\'a}sz}, {Julbe},
  {Karampelas}, {Kewley}, {Klar}, {Kochoska}, {Kohley}, {Kolenberg},
  {Kontizas}, {Kontizas}, {Koposov}, {Kordopatis}, {Kostrzewa-Rutkowska},
  {Koubsky}, {Lambert}, {Lanza}, {Lasne}, {Lavigne}, {Le Fustec}, {Le
  Poncin-Lafitte}, {Lebreton}, {Leccia}, {Leclerc}, {Lecoeur-Taibi},
  {Lenhardt}, {Leroux}, {Liao}, {Licata}, {Lindstr{\o}m}, {Lister}, {Livanou},
  {Lobel}, {L{\'o}pez}, {Managau}, {Mann}, {Mantelet}, {Marchal}, {Marchant},
  {Marconi}, {Marinoni}, {Marschalk{\'o}}, {Marshall}, {Martino}, {Marton},
  {Mary}, {Massari}, {Matijevi{\v{c}}}, {Mazeh}, {McMillan}, {Messina},
  {Michalik}, {Millar}, {Molina}, {Molinaro}, {Moln{\'a}r}, {Montegriffo},
  {Mor}, {Morbidelli}, {Morel}, {Morris}, {Mulone}, {Muraveva}, {Musella},
  {Nelemans}, {Nicastro}, {Noval}, {O'Mullane}, {Ord{\'e}novic},
  {Ord{\'o}{\~n}ez-Blanco}, {Osborne}, {Pagani}, {Pagano}, {Pailler},
  {Palacin}, {Palaversa}, {Panahi}, {Pawlak}, {Piersimoni}, {Pineau}, {Plachy},
  {Plum}, {Poggio}, {Poujoulet}, {Pr{\v{s}}a}, {Pulone}, {Racero}, {Ragaini},
  {Rambaux}, {Ramos-Lerate}, {Regibo}, {Reyl{\'e}}, {Riclet}, {Ripepi}, {Riva},
  {Rivard}, {Rixon}, {Roegiers}, {Roelens}, {Romero-G{\'o}mez}, {Rowell},
  {Royer}, {Ruiz-Dern}, {Sadowski}, {Sagrist{\`a} Sell{\'e}s}, {Sahlmann},
  {Salgado}, {Salguero}, {Sanna}, {Santana-Ros}, {Sarasso}, {Savietto},
  {Schultheis}, {Sciacca}, {Segol}, {Segovia}, {S{\'e}gransan}, {Shih},
  {Siltala}, {Silva}, {Smart}, {Smith}, {Solano}, {Solitro}, {Sordo}, {Soria
  Nieto}, {Souchay}, {Spagna}, {Spoto}, {Stampa}, {Steele},
  {Steidelm{\"u}ller}, {Stephenson}, {Stoev}, {Suess}, {Surdej}, {Szabados},
  {Szegedi-Elek}, {Tapiador}, {Taris}, {Tauran}, {Taylor}, {Teixeira},
  {Terrett}, {Teyssand ier}, {Thuillot}, {Titarenko}, {Torra Clotet}, {Turon},
  {Ulla}, {Utrilla}, {Uzzi}, {Vaillant}, {Valentini}, {Valette}, {van Elteren},
  {Van Hemelryck}, {van Leeuwen}, {Vaschetto}, {Vecchiato}, {Veljanoski},
  {Viala}, {Vicente}, {Vogt}, {von Essen}, {Voss}, {Votruba}, {Voutsinas},
  {Walmsley}, {Weiler}, {Wertz}, {Wevers}, {Wyrzykowski}, {Yoldas},
  {{\v{Z}}erjal}, {Ziaeepour}, {Zorec}, {Zschocke}, {Zucker}, {Zurbach}, \&
  {Zwitter}}]{GaiaCollaboration2018}
{Gaia Collaboration}, {Brown}, A.~G.~A., {Vallenari}, A., {et~al.} 2018, \aap,
  616, A1

\bibitem[{{Helling} \& {Casewell}(2014)}]{helling2014}
{Helling}, C., \& {Casewell}, S. 2014, \aapr, 22, 80

\bibitem[{{Hiranaka} {et~al.}(2016){Hiranaka}, {Cruz}, {Douglas}, {Marley}, \&
  {Baldassare}}]{hiranaka2016}
{Hiranaka}, K., {Cruz}, K.~L., {Douglas}, S.~T., {Marley}, M.~S., \&
  {Baldassare}, V.~F. 2016, ArXiv e-prints, arXiv:1606.09485

\bibitem[{{Karalidi} {et~al.}(2016){Karalidi}, {Apai}, {Marley}, \&
  {Buenzli}}]{karalidi2016}
{Karalidi}, T., {Apai}, D., {Marley}, M.~S., \& {Buenzli}, E. 2016, \apj, 825,
  90

\bibitem[{{K{\"u}mmel} {et~al.}(2011){K{\"u}mmel}, {Kuntschner}, {Walsh}, \&
  {Bushouse}}]{kummel2011}
{K{\"u}mmel}, M., {Kuntschner}, H., {Walsh}, J.~R., \& {Bushouse}, H. 2011,
  ST-ECF Instrument Science Report WFC3-2011-05 (STScI)

\bibitem[{{K{\"u}mmel} {et~al.}(2009){K{\"u}mmel}, {Walsh}, {Pirzkal},
  {Kuntschner}, \& {Pasquali}}]{kummel2009}
{K{\"u}mmel}, M., {Walsh}, J.~R., {Pirzkal}, N., {Kuntschner}, H., \&
  {Pasquali}, A. 2009, \pasp, 121, 59

\bibitem[{{Kuntschner} {et~al.}(2009){Kuntschner}, {Bushouse}, {K{\"u}mmel}, \&
  {Walsh}}]{kuntschner2009}
{Kuntschner}, H., {Bushouse}, H., {K{\"u}mmel}, M., \& {Walsh}, J.~R. 2009,
  ST-ECF Instrument Science Report WFC3-2009-17, 17

\bibitem[{{Kuntschner} {et~al.}(2011){Kuntschner}, {K{\"u}mmel}, {Walsh}, \&
  {Bushouse}}]{kuntschner2011}
{Kuntschner}, H., {K{\"u}mmel}, M., {Walsh}, J.~R., \& {Bushouse}, H. 2011,
  ST-ECF Instrument Science Report WFC3-2011-05, 5

\bibitem[{{Lew} {et~al.}(2016){Lew}, {Apai}, {Zhou}, {Schneider}, {Burgasser},
  {Karalidi}, {Yang}, {Marley}, {Cowan}, {Bedin}, {Metchev}, {Radigan}, \&
  {Lowrance}}]{lew2016}
{Lew}, B.~W.~P., {Apai}, D., {Zhou}, Y., {et~al.} 2016, \apjl, 829, L32

\bibitem[{{Liu} {et~al.}(2016){Liu}, {Dupuy}, \& {Allers}}]{liu2016}
{Liu}, M.~C., {Dupuy}, T.~J., \& {Allers}, K.~N. 2016, \apj, 833, 96

\bibitem[{{Long} {et~al.}(2014){Long}, {Baggett}, {MacKenty}, \&
  {McCullough}}]{long2014}
{Long}, K.~S., {Baggett}, S.~M., {MacKenty}, J.~W., \& {McCullough}, P.~M.
  2014, {Attempts to Mitigate Trapping Effects in Scanned Grism Observations of
  Exoplanet Transits with WFC3/IR}, Tech. rep.

\bibitem[{{Luri} {et~al.}(2018){Luri}, {Brown}, {Sarro}, {Arenou},
  {Bailer-Jones}, {Castro-Ginard}, {de Bruijne}, {Prusti}, {Babusiaux}, \&
  {Delgado}}]{luri2018}
{Luri}, X., {Brown}, A.~G.~A., {Sarro}, L.~M., {et~al.} 2018, \aap, 616, A9

\bibitem[{{Malo} {et~al.}(2013){Malo}, {Doyon}, {Lafreni{\`e}re}, {Artigau},
  {Gagn{\'e}}, {Baron}, \& {Riedel}}]{malo2013}
{Malo}, L., {Doyon}, R., {Lafreni{\`e}re}, D., {et~al.} 2013, \apj, 762, 88

\bibitem[{{Manjavacas} {et~al.}(2018){Manjavacas}, {Apai}, {Zhou}, {Karalidi},
  {Lew}, {Schneider}, {Cowan}, {Metchev}, {Miles-P{\'a}ez}, {Burgasser},
  {Radigan}, {Bedin}, {Lowrance}, \& {Marley}}]{manjavacas2018}
{Manjavacas}, E., {Apai}, D., {Zhou}, Y., {et~al.} 2018, \aj, 155, 11

\bibitem[{{Manjavacas} {et~al.}(2019){Manjavacas}, {Apai}, {Lew}, {Zhou},
  {Schneider}, {Burgasser}, {Karalidi}, {Miles-P{\'a}ez}, {Lowrance}, {Cowan},
  {Bedin}, {Marley}, {Metchev}, \& {Radigan}}]{manjavacas2019b}
{Manjavacas}, E., {Apai}, D., {Lew}, B. W.~P., {et~al.} 2019, \apjl, 875, L15

\bibitem[{{Mann} {et~al.}(2013){Mann}, {Brewer}, {Gaidos}, {L{\'e}pine}, \&
  {Hilton}}]{Mann2013}
{Mann}, A.~W., {Brewer}, J.~M., {Gaidos}, E., {L{\'e}pine}, S., \& {Hilton},
  E.~J. 2013, \aj, 145, 52

\bibitem[{{Marley} {et~al.}(2012){Marley}, {Saumon}, {Cushing}, {Ackerman},
  {Fortney}, \& {Freedman}}]{marley2012}
{Marley}, M.~S., {Saumon}, D., {Cushing}, M., {et~al.} 2012, \apj, 754, 135

\bibitem[{{Marley} {et~al.}(2010){Marley}, {Saumon}, \&
  {Goldblatt}}]{marley2010}
{Marley}, M.~S., {Saumon}, D., \& {Goldblatt}, C. 2010, \apjl, 723, L117

\bibitem[{{Marocco} {et~al.}(2014){Marocco}, {Day-Jones}, {Lucas}, {Jones},
  {Smart}, {Zhang}, {Gomes}, {Burningham}, {Pinfield}, {Raddi}, \&
  {Smith}}]{marocco2014}
{Marocco}, F., {Day-Jones}, A.~C., {Lucas}, P.~W., {et~al.} 2014, \mnras, 439,
  372

\bibitem[{{Metchev} \& {Hillenbrand}(2006)}]{metchev2006}
{Metchev}, S.~A., \& {Hillenbrand}, L.~A. 2006, \apj, 651, 1166

\bibitem[{{Metchev} {et~al.}(2015){Metchev}, {Heinze}, {Apai}, {Flateau},
  {Radigan}, {Burgasser}, {Marley}, {Artigau}, {Plavchan}, \&
  {Goldman}}]{metchev2015}
{Metchev}, S.~A., {Heinze}, A., {Apai}, D., {et~al.} 2015, \apj, 799, 154

\bibitem[{Miles-P{\'{a}}ez {et~al.}(2017)Miles-P{\'{a}}ez, Metchev, Luhman,
  Marengo, \& Hulsebus}]{miles-paez2017}
Miles-P{\'{a}}ez, P.~A., Metchev, S., Luhman, K.~L., Marengo, M., \& Hulsebus,
  A. 2017, \aj, 154, 262.
\newblock \url{https://doi.org/10.3847\%2F1538-3881\%2Faa9711}

\bibitem[{Miles-P{\'{a}}ez {et~al.}(2019)Miles-P{\'{a}}ez, Metchev, Apai, Zhou,
  Manjavacas, Karalidi, Lew, Burgasser, Bedin, Cowan, Lowrance, Marley,
  Radigan, \& Schneider}]{miles-paez2019}
Miles-P{\'{a}}ez, P.~A., Metchev, S., Apai, D., {et~al.} 2019, \apj, 883, 181.
\newblock \url{https://doi.org/10.3847%2F1538-4357\%2Fab3d25}

\bibitem[{{Naud} {et~al.}(2017){Naud}, {Artigau}, {Rowe}, {Doyon}, {Malo},
  {Albert}, {Gagn{\'e}}, \& {Bouchard}}]{naud2017}
{Naud}, M.-E., {Artigau}, {\'E}., {Rowe}, J.~F., {et~al.} 2017, \aj, 154, 138

\bibitem[{{Naud} {et~al.}(2014){Naud}, {Artigau}, {Malo}, {Albert}, {Doyon},
  {Lafreni{\`e}re}, {Gagn{\'e}}, {Saumon}, {Morley}, \& {Allard}}]{naud2014}
{Naud}, M.-E., {Artigau}, {\'E}., {Malo}, L., {et~al.} 2014, \apj, 787, 5

\bibitem[{{Newton} {et~al.}(2014){Newton}, {Charbonneau}, {Irwin},
  {Berta-Thompson}, {Rojas-Ayala}, {Covey}, \& {Lloyd}}]{Newton2014}
{Newton}, E.~R., {Charbonneau}, D., {Irwin}, J., {et~al.} 2014, \aj, 147, 20

\bibitem[{{Norton} {et~al.}(2007){Norton}, {Wheatley}, {West}, {Haswell},
  {Street}, {Collier Cameron}, {Christian}, {Clarkson}, {Enoch}, {Gallaway},
  {Hellier}, {Horne}, {Irwin}, {Kane}, {Lister}, {Nicholas}, {Parley},
  {Pollacco}, {Ryans}, {Skillen}, \& {Wilson}}]{norton2007}
{Norton}, A.~J., {Wheatley}, P.~J., {West}, R.~G., {et~al.} 2007, \aap, 467,
  785

\bibitem[{{Radigan}(2014)}]{radigan2014a}
{Radigan}, J. 2014, \apj, 797, 120

\bibitem[{{Radigan} {et~al.}(2012){Radigan}, {Jayawardhana}, {Lafreni{\`e}re},
  {Artigau}, {Marley}, \& {Saumon}}]{radigan2012}
{Radigan}, J., {Jayawardhana}, R., {Lafreni{\`e}re}, D., {et~al.} 2012, \apj,
  750, 105

\bibitem[{{Radigan} {et~al.}(2014){Radigan}, {Lafreni{\`e}re}, {Jayawardhana},
  \& {Artigau}}]{radigan2014b}
{Radigan}, J., {Lafreni{\`e}re}, D., {Jayawardhana}, R., \& {Artigau}, E. 2014,
  \apj, 793, 75

\bibitem[{{Riaz} {et~al.}(2006){Riaz}, {Gizis}, \& {Harvin}}]{riaz2006}
{Riaz}, B., {Gizis}, J.~E., \& {Harvin}, J. 2006, \aj, 132, 866

\bibitem[{{Robinson} \& {Marley}(2014)}]{robinson2014}
{Robinson}, T.~D., \& {Marley}, M.~S. 2014, \apj, 785, 158

\bibitem[{{Saumon} \& {Marley}(2008)}]{saumon2008}
{Saumon}, D., \& {Marley}, M.~S. 2008, \apj, 689, 1327

\bibitem[{{Schlawin} {et~al.}(2017){Schlawin}, {Burgasser}, {Karalidi},
  {Gizis}, \& {Teske}}]{schlawin2017}
{Schlawin}, E., {Burgasser}, A.~J., {Karalidi}, T., {Gizis}, J.~E., \& {Teske},
  J. 2017, \apj, 849, 163

\bibitem[{{Scholz} {et~al.}(2018){Scholz}, {Moore}, {Jayawardhana}, {Aigrain},
  {Peterson}, \& {Stelzer}}]{scholz2018}
{Scholz}, A., {Moore}, K., {Jayawardhana}, R., {et~al.} 2018, \apj, 859, 153

\bibitem[{{Schwartz} \& {Cowan}(2015)}]{schwartz2015}
{Schwartz}, J.~C., \& {Cowan}, N.~B. 2015, \mnras, 449, 4192

\bibitem[{{Showman} {et~al.}(2018){Showman}, {Tan}, \& {Zhang}}]{showman2018}
{Showman}, A.~P., {Tan}, X., \& {Zhang}, X. 2018, ArXiv e-prints,
  arXiv:1807.08433

\bibitem[{{Smith} {et~al.}(2008){Smith}, {Zavodny}, {Rahmer}, \&
  {Bonati}}]{smith2008}
{Smith}, R.~M., {Zavodny}, M., {Rahmer}, G., \& {Bonati}, M. 2008, in Society
  of Photo-Optical Instrumentation Engineers (SPIE) Conference Series, Vol.
  7021, \procspie, 70210K

\bibitem[{{Snellen} {et~al.}(2014){Snellen}, {Brandl}, {de Kok}, {Brogi},
  {Birkby}, \& {Schwarz}}]{snellen2014}
{Snellen}, I.~A.~G., {Brandl}, B.~R., {de Kok}, R.~J., {et~al.} 2014, \nat,
  509, 63

\bibitem[{{Tan} \& {Showman}(2017)}]{tan2017}
{Tan}, X., \& {Showman}, A.~P. 2017, \apj, 835, 186

\bibitem[{{Tan} \& {Showman}(2018)}]{tan2018}
---. 2018, ArXiv e-prints, arXiv:1809.06467

\bibitem[{{Tremblin} {et~al.}(2016){Tremblin}, {Amundsen}, {Chabrier},
  {Baraffe}, {Drummond}, {Hinkley}, {Mourier}, \& {Venot}}]{tremblin2016}
{Tremblin}, P., {Amundsen}, D.~S., {Chabrier}, G., {et~al.} 2016, \apjl, 817,
  L19

\bibitem[{{Tremblin} {et~al.}(2019){Tremblin}, {Padioleau}, {Phillips},
  {Chabrier}, {Baraffe}, {Fromang}, {Audit}, {Bloch}, {Burgasser}, {Drummond},
  {Gonz{\'a}lez}, {Kestener}, {Kokh}, {Lagage}, \& {Stauffert}}]{tremblin2019}
{Tremblin}, P., {Padioleau}, T., {Phillips}, M.~W., {et~al.} 2019, \apj, 876,
  144

\bibitem[{{Tsuji} \& {Nakajima}(2003)}]{tsuji2003}
{Tsuji}, T., \& {Nakajima}, T. 2003, \apjl, 585, L151

\bibitem[{{Vos} {et~al.}(2017){Vos}, {Allers}, \& {Biller}}]{vos2017}
{Vos}, J.~M., {Allers}, K.~N., \& {Biller}, B.~A. 2017, \apj, 842, 78

\bibitem[{{Wakeford} {et~al.}(2016){Wakeford}, {Sing}, {Evans}, {Deming}, \&
  {Mandell}}]{wakeford2016}
{Wakeford}, H.~R., {Sing}, D.~K., {Evans}, T., {Deming}, D., \& {Mandell}, A.
  2016, \apj, 819, 10

\bibitem[{{West} {et~al.}(2008){West}, {Hawley}, {Bochanski}, {Covey}, {Reid},
  {Dhital}, {Hilton}, \& {Masuda}}]{west2008}
{West}, A.~A., {Hawley}, S.~L., {Bochanski}, J.~J., {et~al.} 2008, \aj, 135,
  785

\bibitem[{{White} \& {Basri}(2003)}]{white2003}
{White}, R.~J., \& {Basri}, G. 2003, \apj, 582, 1109

\bibitem[{{Wright} {et~al.}(2010){Wright}, {Eisenhardt}, {Mainzer}, {Ressler},
  {Cutri}, {Jarrett}, {Kirkpatrick}, {Padgett}, {McMillan}, {Skrutskie},
  {Stanford}, {Cohen}, {Walker}, {Mather}, {Leisawitz}, {Gautier}, {McLean},
  {Benford}, {Lonsdale}, {Blain}, {Mendez}, {Irace}, {Duval}, {Liu}, {Royer},
  {Heinrichsen}, {Howard}, {Shannon}, {Kendall}, {Walsh}, {Larsen}, {Cardon},
  {Schick}, {Schwalm}, {Abid}, {Fabinsky}, {Naes}, \& {Tsai}}]{wright2010}
{Wright}, E.~L., {Eisenhardt}, P.~R.~M., {Mainzer}, A.~K., {et~al.} 2010, \aj,
  140, 1868

\bibitem[{{Yang} {et~al.}(2015){Yang}, {Apai}, {Marley}, {Saumon}, {Morley},
  {Buenzli}, {Artigau}, {Radigan}, {Metchev}, {Burgasser}, {Mohanty},
  {Lowrance}, {Showman}, {Karalidi}, {Flateau}, \& {Heinze}}]{yang2015}
{Yang}, H., {Apai}, D., {Marley}, M.~S., {et~al.} 2015, \apjl, 798, L13

\bibitem[{{Yang} {et~al.}(2016){Yang}, {Apai}, {Marley}, {Karalidi}, {Flateau},
  {Showman}, {Metchev}, {Buenzli}, {Radigan}, {Artigau}, {Lowrance}, \&
  {Burgasser}}]{yang2016}
---. 2016, \apj, 826, 8

\bibitem[{{Zhou} {et~al.}(2017){Zhou}, {Apai}, {Lew}, \&
  {Schneider}}]{zhou2017}
{Zhou}, Y., {Apai}, D., {Lew}, B.~W.~P., \& {Schneider}, G. 2017, ArXiv
  e-prints, arXiv:1703.01301

\bibitem[{{Zhou} {et~al.}(2016){Zhou}, {Apai}, {Schneider}, {Marley}, \&
  {Showman}}]{zhou2016}
{Zhou}, Y., {Apai}, D., {Schneider}, G.~H., {Marley}, M.~S., \& {Showman},
  A.~P. 2016, \apj, 818, 176

\bibitem[{{Zhou} {et~al.}(2018){Zhou}, {Apai}, {Metchev}, {Lew}, {Schneider},
  {Marley}, {Karalidi}, {Manjavacas}, {Bedin}, {Cowan}, {Miles-P{\'a}ez},
  {Lowrance}, {Radigan}, \& {Burgasser}}]{zhou2018}
{Zhou}, Y., {Apai}, D., {Metchev}, S., {et~al.} 2018, \aj, 155, 132

\bibitem[{{Zhou} {et~al.}(2019){Zhou}, {Apai}, {Lew}, {Schneider},
  {Manjavacas}, {Bedin}, {Cowan}, {Marley}, {Radigan}, {Karalidi}, {Lowrance},
  {Miles-P{\'a}ez}, {Metchev}, \& {Burgasser}}]{zhou2019}
{Zhou}, Y., {Apai}, D., {Lew}, B. W.~P., {et~al.} 2019, \aj, 157, 128

\bibitem[{{Zuckerman} {et~al.}(2004){Zuckerman}, {Song}, \&
  {Bessell}}]{zuckerman2004}
{Zuckerman}, B., {Song}, I., \& {Bessell}, M.~S. 2004, \apj, 613, L65

\end{thebibliography}
\appendix
\restartappendixnumbering 

\section{Contamination Model}
To provide a quantitative estimate of the contamination, we use a model with three 1-D Moffat profile to fit the horizontally summed (sum of pixels of column 440-570 in Figure \ref{fig:fov}b, corresponding to the $1.1-1.7\rm\, \mu m$ region of GU\,Psc\,b's spectrum) pixel count rates of the GU\,Psc\,b, the nearby galaxy, and the reference star. 
The reduced chi-square from the model fitting is large ($\sim$ 700) because of the significant deviation at the wing. Based on the best-fit Moffat profiles, the galaxy's and the reference star's flux in the eight-pixel wide shaded region is about 10\% and 3\% of the GU\,Psc\,b's flux respectively.
Contamination levels in the $J'$- and $H'$-bands are thus lower than 13\% because GU\,Psc\,b's spectral intensities are higher in these bands than the averaged intensity over 1.1-1.7$\rm\, \mu m$. 
 As mentioned in Section \ref{sec:contam}, the low contamination level ($<13\%$) and the measured variability of the galaxy and reference star together suggests that the detected flux variation of GU\,Psc\,b is intrinsic. 
\begin{figure}[hbtp]
    \begin{minipage}[c]{0.4\textwidth}
    \caption{Top: The fitting result of the 3-Moffat model (solid red line) to the horizontally summed (i.e. sum of pixel count rates from the Column 440-570 in Figure \ref{fig:fov}) pixel count rates (dashed blue line). The three Moffat profiles are plotted in orange dashed, dotted, and solid lines. Middle panel: Same as the top panel with zoomed-in view of the fitting result for the Galaxy. Bottom panel: The residual between the model and measured count rate $\Delta C$ in unit of the observation noise $\epsilon_c$ (photon and readout noise). }
    \label{fig:contam}
    \end{minipage}\hfill
      \begin{minipage}[c]{0.7\textwidth}
      \centering
        \includegraphics[width=0.8\textwidth]{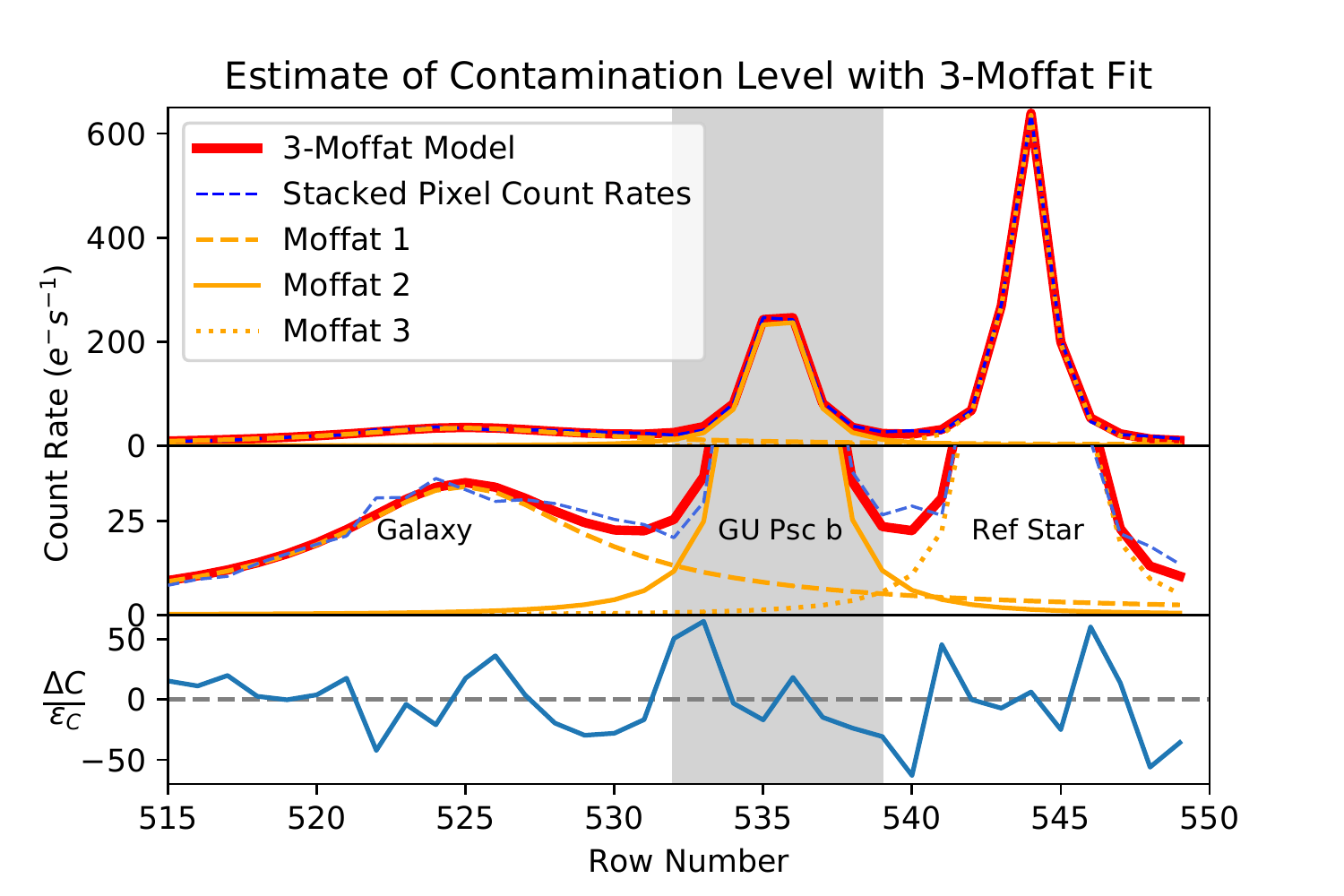}
        \end{minipage}
\end{figure}

\begin{figure}[hbtp]
\begin{minipage}[c]{0.4\textwidth}
\caption{A graphic illustration of the difference between Ordinary Distance Least Square Regression and Orthogonal Distance Least Square Regression. In brief, the latter minimizes the uncertainty-weighted \textit{orthogonal} distance between data and model while the prior minimizes the uncertainty-weighted \textit{vertical} distance between data and model.}    \label{fig:odr}
\end{minipage}\hfill
      \begin{minipage}[c]{0.7\textwidth}
      \centering
    \includegraphics[width=.67\textwidth]{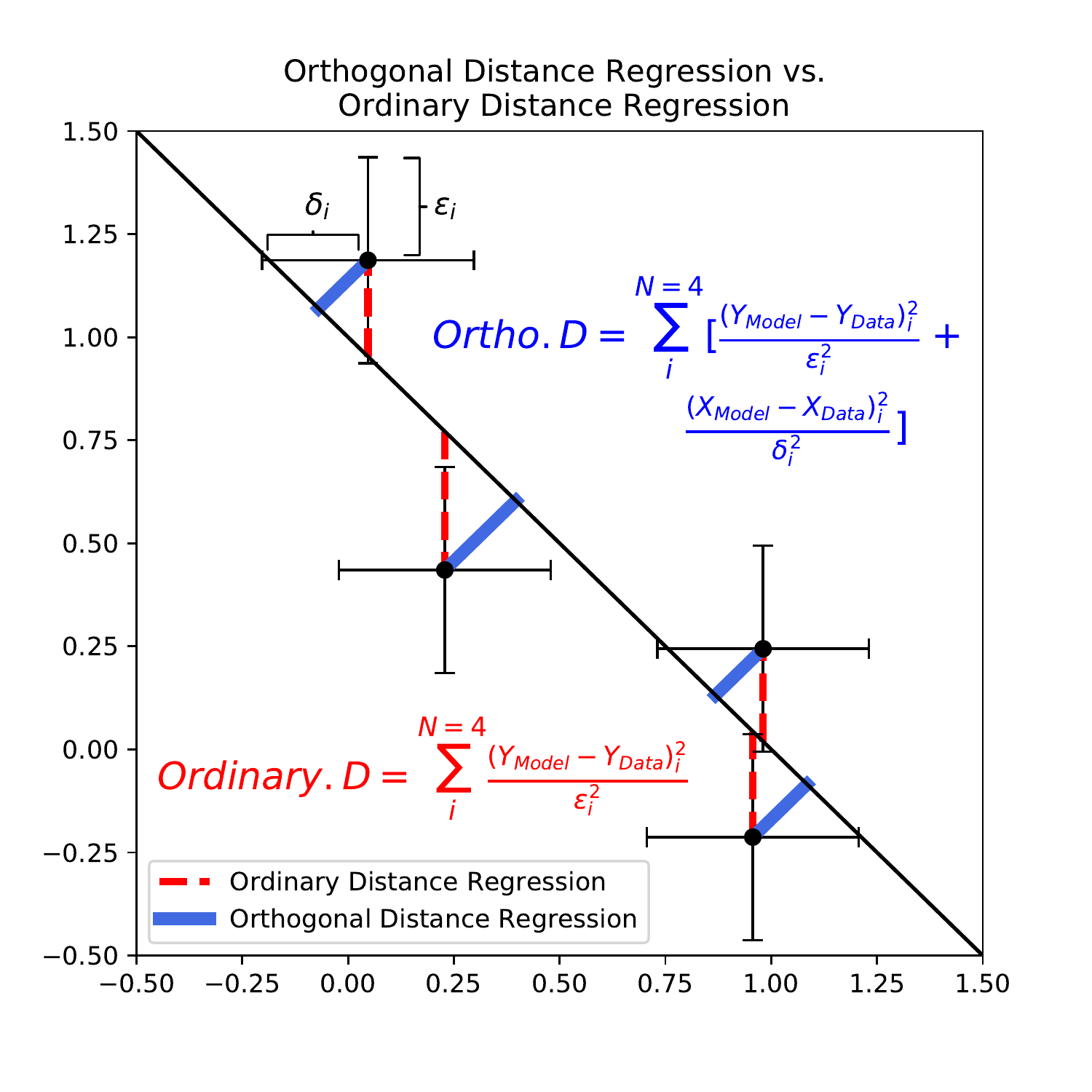}
  \end{minipage}
\end{figure}

\begin{figure}
    \centering
    \includegraphics[width=0.8\textwidth]{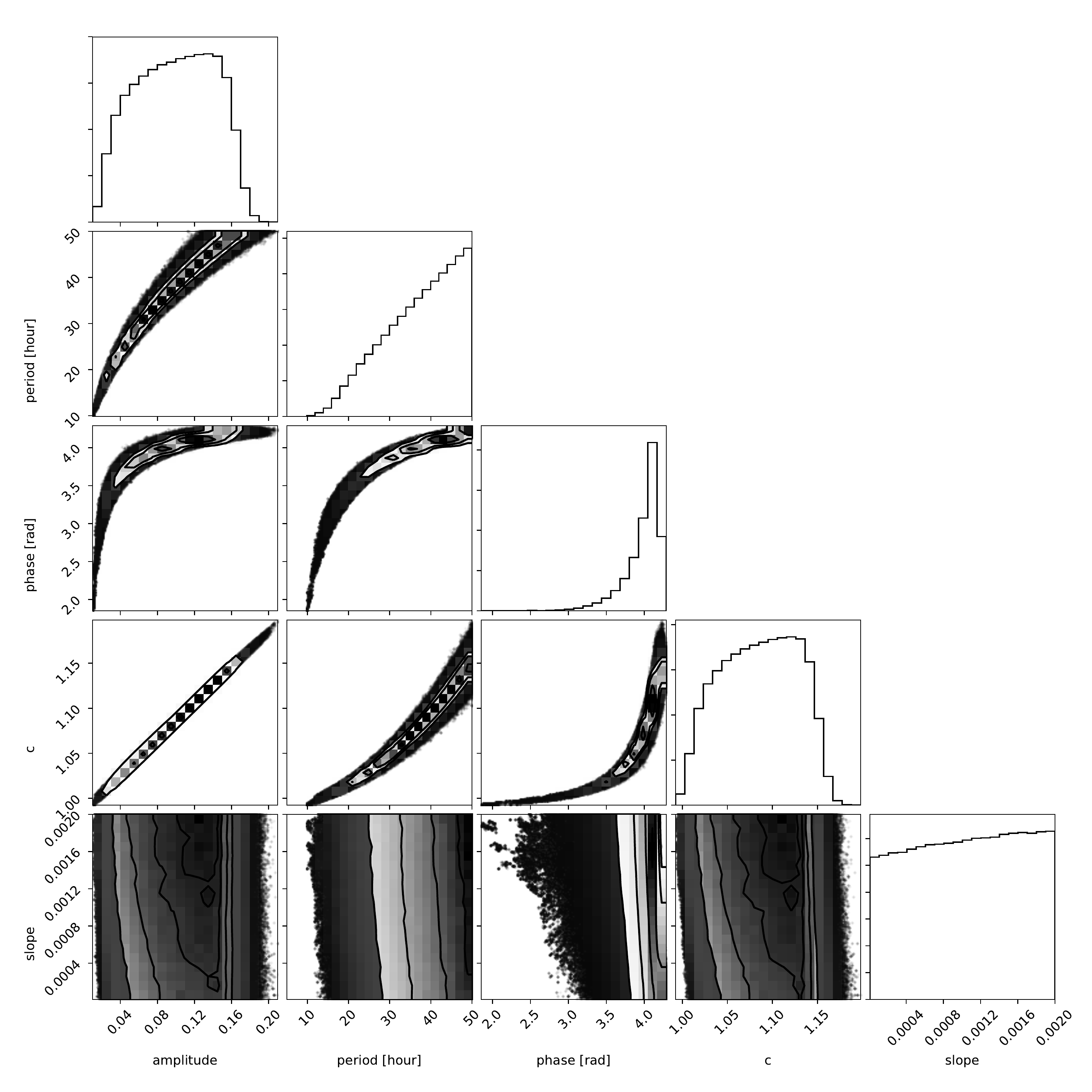}
    \caption{The posterior distribution results from Markov Chain Monte Carlo (MCMC) Method with \texttt{emcee} for a simple sine wave model on top of a linear slope. We use log-uniform priors of period P=[5, 50 hours], phase= [0, 2$\pi$], baseline c= [0.5, 1.5], amplitude = [0.003, 1.2], and slope = [$10^{-5},2\times 10^{-3}$ $\rm hour^{-1}$]. We then run MCMC with 50 walkers for 500,000 steps. \edit1{We note that the upper bound (50 hours) of period posterior distribution is equivalent to the upper bound of prior. Therefore, the upper bounds of the period, amplitude, and baseline are unconstrained based on the MCMC result.} The posterior distribution shows that including the possible visit-long slope (c.f., Section \ref{sec:contam}) as a free parameter does not affect the result.}
    \label{fig:mcmc}
\end{figure}

\end{document}